\documentclass[gmd,manuscript]{copernicus}

\usepackage{tabularx}
\usepackage{multirow}
\usepackage{algorithmic}
\usepackage{algorithm}
\usepackage{amsthm}
\usepackage{float}
\usepackage{subfig}

\usepackage{orcidlink}

\begin{document}
\nolinenumbers
\title{Spatialize v1.0: A Python/C++ Library for Ensemble Spatial Interpolation}


\Author[1,2][\href{mailto:felipe.navarro@amtc.cl}{felipe.navarro@amtc.cl}]{Felipe Navarro~\orcidlink{0000-0002-6951-3504}}{}  
\Author[1,2]{Alvaro F. Egaña~\orcidlink{0000-0001-8720-4783}}{} 
\Author[1,2]{Alejandro Ehrenfeld~\orcidlink{0000-0001-5370-0172}}{}
\Author[1,2]{Felipe Garrido}{}
\Author[1,2]{María Jesús Valenzuela}{}
\Author[3]{Juan F. Sánchez-Pérez~\orcidlink{0000-0003-3382-8165}}{}

\affil[1]{Advanced Mining Technology Center (AMTC)\\
  Universidad de Chile\\
  Avenida Tupper 2007, RM, Chile\\
  URL: \url{http://www.amtc.cl}  }
\affil[2]{Advanced Laboratory for Geostatistical Supercomputing (ALGES)\\
  Department of Mining Engineering - AMTC\\
  Faculty of Mathematical and Physical Sciences\\
  Universidad de Chile\\
  Avenida Tupper 2069, RM, Chile\\
  URL: \url{http://www.alges.cl}  }
\affil[3]{Department of Applied Physics and Naval Technology, Universidad Politécnica de Cartagena (UPCT), Cartagena, Spain}




\runningtitle{Spatialize: A Library for Ensemble Spatial Interpolation}

\runningauthor{Navarro et al.}

\received{}
\pubdiscuss{} 
\revised{}
\accepted{}
\published{}


\firstpage{1}

\maketitle

\begin{abstract}
In this paper, we present Spatialize, an open-source library that implements \emph{ensemble spatial interpolation}, a novel method that combines the simplicity of basic interpolation methods with the power of classical geostatistical tools, like Kriging. It leverages the richness of stochastic modelling and ensemble learning, making it robust, scalable and suitable for large datasets. In addition, Spatialize provides a powerful framework for uncertainty quantification, offering both point estimates and empirical posterior distributions. It is implemented in Python 3.x, with a C++ core for improved performance, and is designed to be easy to use, requiring minimal user intervention. This library aims to bridge the gap between expert and non-expert users of geostatistics by providing automated tools that rival traditional geostatistical methods. Here, we present a detailed description of Spatialize along with a wealth of examples of its use.
\end{abstract}


\introduction  

A significant challenge in the field of geosciences is the issue of sparsity that is often observed in spatial databases, such as soil properties, climate data, or mineral concentrations, which are characterised at limited point locations. \citep{Li2014SpatialReview}. The presence of these data gaps hinders a comprehensive understanding of the variable's domain. The central issue, therefore, is estimating values at unmeasured locations. Various spatial interpolation algorithms have been developed for this purpose.

Geostatistics is a field focused on the analysis, estimation, and modelling of spatial variables. Unlike traditional statistics, geostatistics emphasises the spatial dependencies between observations \citep{MAROUFPOOR2020229}. 
The Kriging technique, the most relevant exponent of geostatistical interpolation \citep{McKinley:2020, Kirkwood:2022}, was initially devised for the estimation of gold reserves \citep{Kleijnen:2017, VirdeeKottegoda:1984}. As an unbiased linear estimator that minimises estimation error at each position, Kriging is commonly referred to as BLUE (Best Linear Unbiased Estimator) \citep{McKinley:2020, FischerGetis:2009, Varouchakis:2012, Abzalov:2016}. Beyond providing robust estimates, Kriging also facilitates the calculation of estimation variance, which is widely used for assessing spatial prediction uncertainty \citep{Abzalov:2016, Varouchakis:2012}.
A variety of user-friendly tools are available for the implementation of Kriging, including \href{https://pykrige.readthedocs.io/en/latest/}{PyKrige}, \href{https://pysal.org/}{PySAL}, \href{https://cran.r-project.org/web/packages/gstat/gstat.pdf}{gstat}, \href{https://cran.r-project.org/web/packages/automap/automap.pdf}{automap}, \href{https://cran.r-project.org/web/packages/geoR/geoR.pdf}{geoR}, and \href{https://cran.r-project.org/web/packages/fields/fields.pdf}{fields}. 
Nevertheless, it should be noted that the use of Kriging without actual knowledge of the model may result in suboptimal and misleading outcomes \citep{OLIVER201456, Assibey-Bonsu:2017}. In particular, parameter selection and spatial continuity modelling have a significant effect on the accuracy of Kriging estimates \citep{Abzalov:2016, Chiles2018, Pannecoucke2020}.
However, the correct determination of these inputs requires substantial expertise and data \citep{FischerGetis:2009, WANG:2017, Pannecoucke2020, deSousa:2020}, which creates a significant barrier for most potential users. 
Moreover, the task becomes increasingly complex when the variables under study are of a dynamic, spatio-temporal nature \citep{Samson:2022, Boroh2022}, or are not structurable as a regular grid \citep{OLIVER201456}.

In summary, spatial interpolation tasks, when assessed from the perspective of classical geostatistical analysis, can be time-consuming and require considerable expertise. Consequently, there is a need for more straightforward yet effective spatial interpolation methods that can address highly dynamic spatial problems without necessitating manual spatial analysis tasks.

In contrast to geostatistical approaches, deterministic models employ straightforward calculations; nevertheless, they are only capable of producing estimations \citep{Li2014SpatialReview}. 
The most widely applied of these methods is inverse distance weighting (IDW), a simple yet powerful spatial interpolation method that uses a weighted average of surrounding point values to estimate the unknown value at an unsampled location \citep{Mitas1988GeneralProblem}.
In recent years, variants of IDW have been successfully used in a variety of applications, including estimation of air pollution levels \citep{LIJialin2017}, soil moisture \citep{Abdulwadood2021} and water quality \citep{KHOUNI2021101892}. The main limitation of IDW is that it does not take into account the spatial structure or correlation of the variable being interpolated. This can lead to over-smoothing or under-smoothing of the estimated values, depending on the degree of spatial correlation in the data \citep{Li_2021IDWLimitations}. 

Another promising approach for spatial interpolation is the use of machine learning-based methods, which can learn complex spatial relationships from large datasets without requiring manual spatial modelling \citep{LiJinHeap2011, Kirkwood2022}. For example, \citet{LeirvikThomasMeng2021} and \citet{Benfeng2019} proposed deep learning-based spatial interpolation methods to estimate solar radiation and interpolate seismic data, respectively. 
Nevertheless, challenges that arise from deep learning models are, firstly, the need for large amounts of data and computational resources to train them, and secondly, the necessity to measure additional variables other than the one under study. This is especially true for complex spatial-temporal problems, where the number of input variables and temporal observations can be substantial \citep{Hamdi2022SpatioTemDataMining}. 
A further challenge in using deep learning for spatial interpolation is the difficulty of interpreting the results. These models are often referred to as `black boxes', meaning that the process by which predictions are derived remains uncertain. This can be problematic in situations where transparency and interpretability are important, such as in environmental applications \citep{PAUDEL_2023DeepLearningInterpret, Qingmin_2013RegressionKriging, Susanto_2016SpatiotemporalInterpolation}.

In addressing the need for a simple and flexible spatial interpolation technique, able to adapt to highly dynamic phenomena, scalable to big data, interpretable, and most importantly widely accessible to the entire geoscientific community, \citet{Menafoglio:2018} and \citet{Egana2021ESI} independently proposed a new state-of-the-art spatial interpolation method based on ensemble learning. This method combines the simplicity of methods such as IDW with the power of Kriging spatial analysis, which the authors of the latter named Ensemble Spatial Interpolation (ESI).
This model is able to provide reliable estimates that are comparable to those of Kriging, while eliminating the need for manual spatial continuity modelling.
Its main features are: (a) it is based on a stochastic space partitioning process, which aids in managing large datasets; (b) it is built under an ensemble scheme, which guarantees robustness despite the use of weak local interpolation functions with small subsets of data, and (c) it provides a powerful framework for uncertainty quantification, as it is based on a Bayesian scheme, thus yielding an empirical posterior distribution of the estimate (instead of a single point estimate).

The aim of this article is to present \verb|spatialize|, a novel software library that facilitates an efficient implementation of ESI. \verb|Spatialize| has been designed to be easy to use, efficient and flexible. The core of the library is implemented in C++ with a Python 3.x programming API. It is available as an open-source project, making it accessible to researchers and practitioners in industry and academia.
The subsequent sections provide a comprehensive overview of the ESI model and the Spatialize library, including its features and capabilities. We also present several examples of how the library can be used in practical applications. Finally, the future directions of the library and its potential impact on spatial interpolation research and practice will be discussed.

\section{Ensemble spatial interpolation}\label{ESI_sec}

Ensemble learning is usually regarded as the statistical and computational conception of the ``wisdom of the crowd", whose idea is to collect and combine the points of view of many experts to produce an ensemble result \citep{Egana2021ESI}. An ensemble model $\hat{z}$ can be formulated as:
\begin{equation}\label{ag_func}
    \hat{z} = G(f_1(x^*), \cdots, f_m(x^*))
\end{equation} \smallskip

Where $x^*$ is a vector of covariances and $\{f_1, \cdots, f_m\}$ is a set of weak voter (regression, classification, interpolation, etc.) functions. Function $G$ is an aggregation function that combines the responses from each voter function and it can be as simple as majority voting for classification, \citep{friedman2000additive,collins2002logistic, Dzeroski2004IsOne, Hothorn2005BundlingTrees, Reid2009RegularizedGeneralization}, averaging for regression, or more sophisticated approaches such as a mixture of experts (MoE) \citep{Jacobs1991AdaptiveExperts, Jacobs1995MethodsAssessments, Jordan1995ConvergenceArchitectures, Cohen2000AArchitecture}.

\subsection{Weak voter function set generation}
\label{weak_voter}

\begin{figure*}[htb]
    \centering
    \includegraphics[width=1.0\columnwidth]{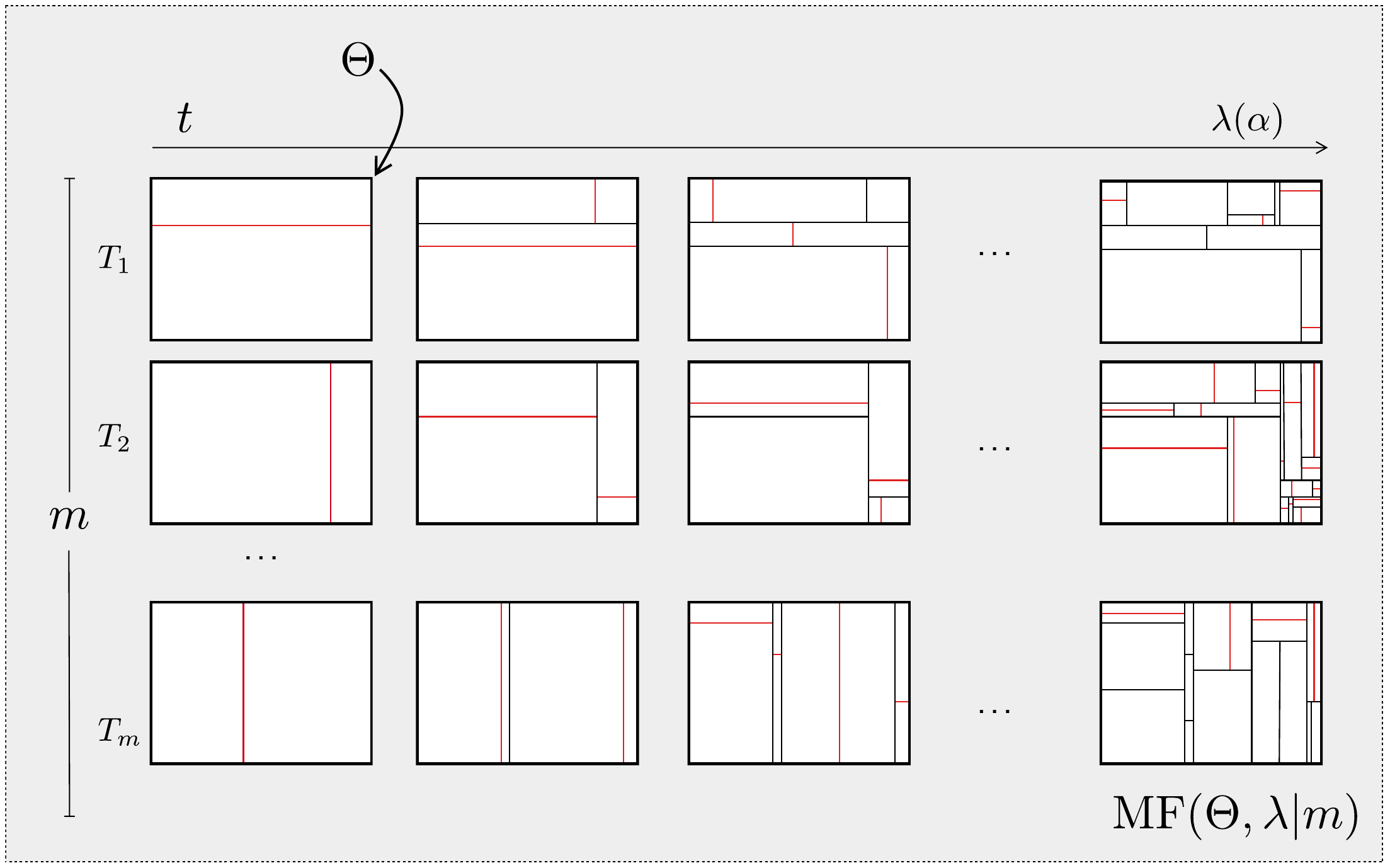}
    \caption{Generative process to draw a time-dependent stochastic partition set of size $m$. Time goes along the x-axis, ending at time given by $\lambda(\alpha)$. Red lines indicate the current time partition cut.}
    \label{fig:generative_process}
\end{figure*}

\begin{figure*}[htb]
    \centering
    \includegraphics[width=1.0\columnwidth]{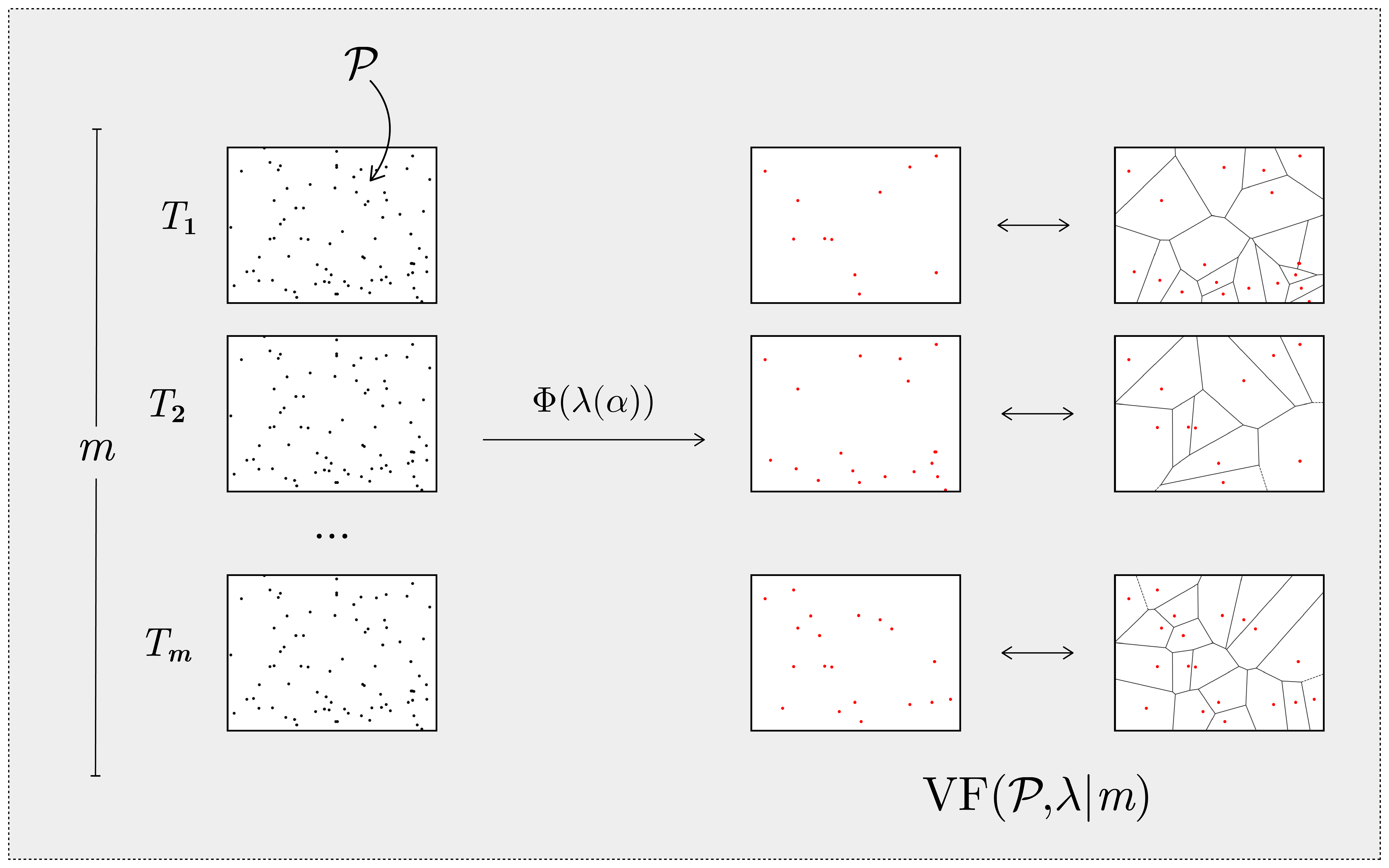}
    \caption{Generative process to draw a topological stochastic partition set of size $m$. The number of Voronoi nuclei is given by $\lambda(\alpha)$. Red points indicate the selected locations, among the sampled locations, for the nuclei.}
    \label{fig:generative_process_voronoi}
\end{figure*}

In the ESI model, the construction of the set of weak voter functions is achieved through the concept of ``bootstrapping the space" \citep{Egana2021ESI}, which involves generating different spatial configurations through random partitioning of the space where the data are located. These partitions are generated in a way that any combination of data is possible, while preserving their spatial locations and avoiding data clustering (Figure \ref{fig:generative_process}).
Each partition creates unique data subsets within the partition cells, where any spatial interpolation method can be applied. An unmeasured location is then estimated by combining (aggregating) the estimates derived from all the partition elements across the set of partitions where that location falls \citep{Egana2021ESI}.

It is well known that a spatial partition data structure can be represented as a tree \citep{samet1984quadtree}, where nodes represent partition spaces and edges indicate containment relationships. This representation enables efficient spatial data querying and operations on the data contained in the spaces, similar to how certain related hierarchical data structures, such as k-d trees and octrees, are used in spatial indexing. Thus, in practice, generating a set of random partitions is equivalent to generating a forest of random tree structures, analogous to how multiple decision trees form a random forest.

The Spatialize library uses two methodologies to generate these partitions of space: \\

\textbf{a) Mondrian Forests (MF):} As proposed by \citet{Egana2021ESI} and introduced by \citet{Lakshminarayanan2014MondrianForests}. The latter is a non-parametric Bayesian strategy that is employed for both classification and online regression, which has been demonstrated to be as efficacious as other high-level ensemble learning methods, such as random forests \citep{Breiman2001RandomForests} or additive trees \citep{Hastie2009ElementsEd}.

In the context of ESI, MF is a collection of random tree structures defined as:
\begin{equation}
   \text{MF} (\Theta, \lambda|m) = \{T_1, \cdots, T_m\}, \: \: T_k \sim \text{MT} (\Theta, \lambda)
\end{equation} 

Where $m$ is the number of random tree structures in the forest and $\text{MT} (\Theta, \lambda)$ is the generative process, described in Algorithm \ref{alg:alg1}, which produces random samples of tree structures. The latter represents random space partitions of the target domain $\Theta$ that are assumed to be 
$\Theta = [a_1, b_1] \times \cdots \times [a_d, b_d] \subset \mathbb{R}^d$.

\begin{algorithm}
\caption{$\text{MT} (\Theta, \lambda)$ Sampling Algorithm}
\label{alg:alg1}

\begin{algorithmic}[1]
    \STATE \textbf{procedure} Main()
        \STATE SampleMT($\Theta, \lambda$)  \COMMENT{$\Theta = [a_1, b_1] \times \cdots \times [a_d, b_d]$}
    \STATE \textbf{end procedure}
    
    \STATE \textbf{procedure} SampleMT($\theta, \lambda$)
        \STATE SampleMTBranch($\theta, \lambda, 0$)
    \STATE \textbf{end procedure}
    
    \STATE \textbf{procedure} SampleMTBranch($\theta, \lambda, \tau$)
        \STATE $E \sim \textrm{Exp}(\mu(\theta))$   \COMMENT{$\mu$ is a measure on $\mathbb{R}^d$} \label{alg:alg1:exp}
        \IF{$(\tau + E) < \lambda$}  \label{alg:alg1:time} 
            \STATE $d_x \sim \textrm{Discrete}(p_1, \ldots, p_d)$ \label{alg:alg1:disc}
            \STATE $x \sim U([a_{d_x}, b_{d_x}])$ \label{alg:alg1:unif}
            \STATE SampleMTBranch($\theta^>, \lambda, (\tau + E)$)
            \STATE SampleMTBranch($\theta^<, \lambda, (\tau + E)$)
        \ENDIF
    \STATE \textbf{end procedure}
\end{algorithmic}
\end{algorithm}

Algorithm \ref{alg:alg1} implements a temporal stochastic process that generates nested random partitions aligned with coordinate axes. 
Parameter $\lambda$ represents the process's finite lifetime, while $\tau$ represents the elapsed time through recursion levels.
The recursive partitioning process is as follows: 
Line \ref{alg:alg1:exp} samples the `time' until the next cut in the sub-box $/theta$ from an exponential distribution, parametrized such that $\mathbb{E}(E) = 1 / \mu(\theta)$, where $\mu(\theta) = \sum_{i=1}^d (b_i^{\theta} - a_i^{\theta})$. This ensures smaller sub-boxes are less likely to be partitioned.
In line \ref{alg:alg1:disc}, the dimension to be partitioned is selected, where $p_k$ determines the probability of selecting dimension $k$. For a sub-box $\theta$, $p_k$ is proportional to $(b_k^{\theta} - a_k^{\theta})$, favouring the partition of larger sides.
In line \ref{alg:alg1:unif}, the cut point is randomly determined along the selected dimension, creating sub-boxes $\theta^>$ and $\theta^<$.
Figure \ref{fig:generative_process} illustrates the partitioning process, with each red line representing a cut at time $t$ for all $T_k$.  \smallskip

\textbf{b) Voronoi Forests (VF):} A variation of a) which employs Voronoi partitions instead of Mondrian trees, in a manner analogous to that described by \citet{Menafoglio2018RandomDomains}. 
However, rather than using a fixed number of nuclei, Spatialize employs a random number per partition, calibrated to ensure that the expected number of data points per cell matches that of a Mondrian tree.

Thus, a Voronoi Forest (VF) is a collection of structures defined as:
\begin{equation}
   \text{VF} (\Theta, \lambda | m) = \{T_1, \cdots, T_m\}, \quad T_k \sim \text{VT} (\Theta, \lambda)
\end{equation}

where $\text{VT} (\Theta, \lambda)$ is the generative process that produces a Voronoi partition.
The process begins with the selection of $K$, the number of Voronoi nuclei, which is drawn from a Poisson distribution with parameter $\lambda$.
Next, a random sample of $K$ nuclei, denoted by $\Phi_K = \{c_1,...,c_K\} \subseteq \Theta$, is randomly generated.
Finally, the target domain $\Theta$ is partitioned by assigning all contained locations $x \in \Theta$ to the nearest nuclei based on the Euclidean distance. A Voronoi cell is thus defined by $\mathcal{L}_i = \{ x \in \Theta : \| x - c_i \| \leq \| x - c_j \| \ \forall c_j \in \Phi_K,  j \neq i \}$. The process of generating a Voronoi Forest is illustrated in Figure \ref{fig:generative_process_voronoi}.

\subsubsection{Model training}
\label{sec:model_training}

Let us define \textit{conditioning data} as a set $\mathcal{M} = \{ z_j \}_{j=1}^{N_s}$ of $N_s$ measurements of a variable of interest, obtained at specific spatial locations $\mathcal{P} = \{ \textbf{x}_j \}_{j=1}^{N_s}$ within a particular region of a $d$-dimensional space. 
The classical formulation of spatial interpolation can be stated as: find a $d$-variate function $\textbf{S}_{(\mathcal{P}, \mathcal{M})}$ that fulfils the condition  $\textbf{S}_{(\mathcal{P}, \mathcal{M})}(\textbf{x}_j) = z_j$, $j = \{1, \cdots, N_s\}$ \footnote{The sub-index $({\mathcal{P}, \mathcal{M}})$ indicates that the interpolation function is constructed using both the values of the measurements and their locations.}. 

Now, let us assume $\mathcal{P} \subset \Theta$. Then, both a Mondrian Forest, $\text{MF} (\Theta, \lambda| m)$, and a Voronoi Forest, $\text{VF} (\Theta, \lambda| m)$, can be `trained' when a set of $d$-dimensional data points (the conditioning data) are used to condition the sampling of $\{T_1, \cdots, T_m\}$. We see, then, that: \\

\textbf{a) When using Mondrian Forests}

A trained Mondrian Forest is defined as:
\begin{equation}\label{eq:TMF}
   \text{MF} (\Theta, \lambda| \mathcal{P}, m) = \{T_1, \cdots, T_m\}, \: \: T_k \sim \text{MT} (\Theta, \lambda| \mathcal{P})
\end{equation}

\begin{figure*}[htb]
    \label{fig:mondrian_tree}
    \centering
    \subfloat[Mondrian tree\label{fig:mondrian_tree:a}] {{\includegraphics[width=0.5\columnwidth]{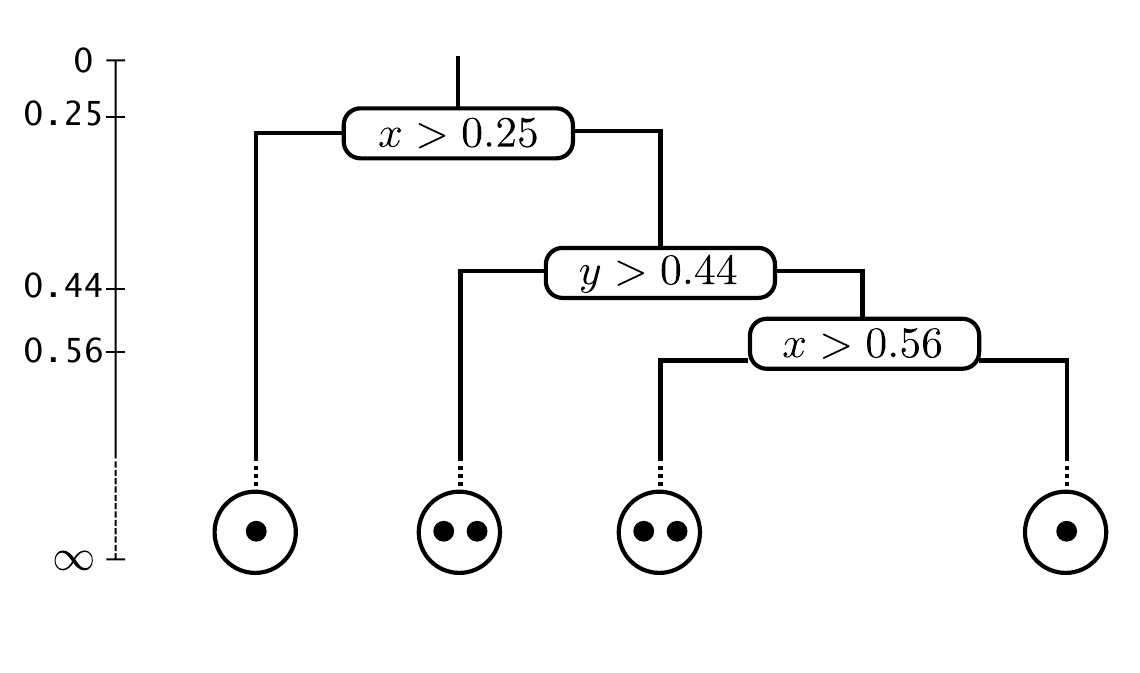} }}
    \subfloat[Conditioning data\label{fig:mondrian_tree:b}] {{\includegraphics[width=0.3\columnwidth]{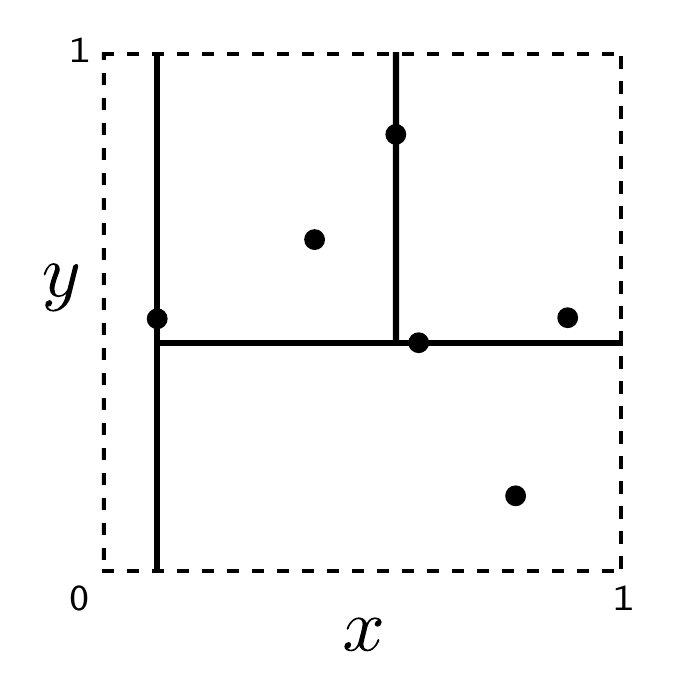} }}
    
    \caption{Tree structure schema. (a) decision tree structure. (b) The spatial partition corresponding to the decision tree structure shown in (a). Black points represent conditioning data.}
\end{figure*}

\begin{figure}
    \centering
    \includegraphics[width=0.8\columnwidth]{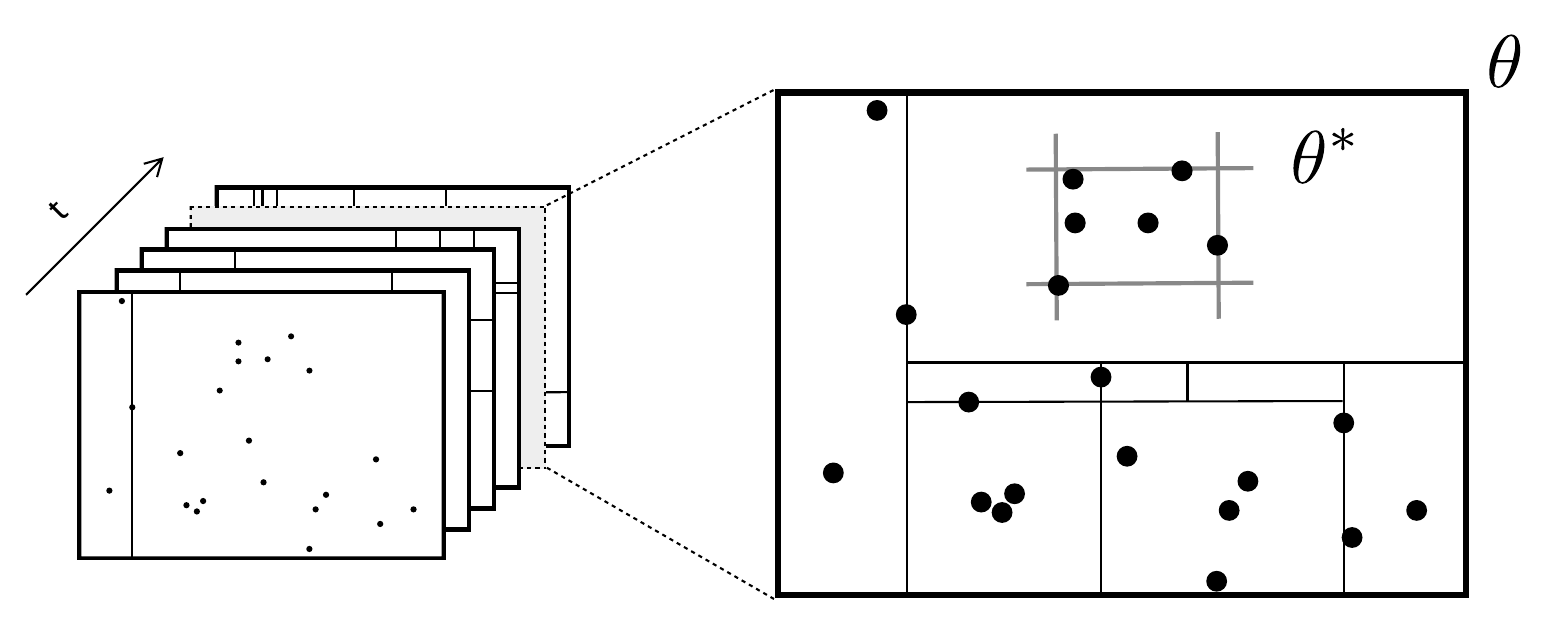}
    \caption{Illustration of $\theta^*$ given a set of data points, which enables Algorithm \ref{alg:alg1} to be trained using sample data.}
    \label{fig:teta_asterisco}
\end{figure}

A Mondrian tree can be trained by conditioning the partitioning process to the data (Figure \ref{fig:mondrian_tree:b}).
Thus, a trained random partition set, $\text{MF} (\Theta, \lambda| \mathcal{P}, m)$, can be obtained by modifying Algorithm \ref{alg:alg1} as follows: 
\begin{itemize}
    \item For any box $\theta$ define $\theta^* = k(\theta , \mathcal{P})$ as the smallest sub-box containing all conditioning positions in $\theta$ (see Figure \ref{fig:teta_asterisco}).
    \item The probability of splitting a sub-box (line \ref{alg:alg1:exp}) is replaced by: Sample $E \sim Exp(\mu(\theta^*))$.
    \item Lines \ref{alg:alg1:disc} and \ref{alg:alg1:unif} in Algorithm \ref{alg:alg1} are adjusted to work on $\theta^*$ instead of $\theta$.
\end{itemize}

As a result of this modification, sub-boxes containing highly concentrated data are more likely to be partitioned. This offers the advantage of ensuring that most leaf nodes (i.e. the resulting sub-boxes) contain a reasonable amount of conditioning data, thus avoiding spatial clustering. \\

\textbf{b) When using Voronoi Forests}

A trained Voronoi Forest (VF) is defined as:
\begin{equation} \label{eq:TVF}
   \text{VF} (\Theta, \lambda |\mathcal{P}, m) = \{T_1, \cdots, T_m\}, \quad T_k \sim \text{VT} (\Theta, \lambda | \mathcal{P})
\end{equation} \smallskip

A trained random partition set, $\text{VF} (\Theta, \lambda| \mathcal{P}, m)$, can be obtained 
by sampling the $m$ sets of Voronoi nuclei from the locations at which measurements of the variable of interest are available, as opposed to being sampled from $\Theta$. For each tree, this results in the set $\Phi_K = \{c_1, \cdots ,c_K\}$ where $K \leq N_s$ and $ \Phi_K \subseteq \mathcal{P}$. 

Thus, the process to obtain samples from $\text{VT} (\Theta, \lambda| \mathcal{P})$ is as follows:
\begin{itemize}
    \item Sample $K \sim Poisson(\lambda),  1 \leq K \leq N_s$.
    \item Sample $\Phi_K = \{c_1,...,c_K\}$ from the measured locations $\mathcal{P}$ 
    \item Establish each Voronoi cell as $\mathcal{L}_i = \{ x \in \Theta : \| x - c_i \| \leq \| x - c_j \| \ \forall c_j \in \Phi_K,  j \neq i \}$.
\end{itemize}

Sampling from the measured locations $\mathcal{P}$ ensures that all partition cells $\mathcal{L}_1, \cdots, \mathcal{L}_K$, generated by sampling $\text{VT} (\Theta, \lambda| \mathcal{P})$, will contain at least one measured location.

\subsubsection{Weak voter function set}
For an unmeasured position $x^* \in \Theta$, we define $\mathcal{L}_k \subset (\mathcal{P}, \mathcal{M})$ as the set of conditioning data points contained within the partition cell where $x^*$ falls into in tree $T_k$. \smallskip

Then, let us consider a base interpolation function $\textbf{S}_{(\mathcal{P}, \mathcal{M})}$, which can be any spatial interpolator for which no additional information, other than measurements and their locations, is required to interpolate new positions -- such as Kriging or IDW. 
 
Now, for each tree $T_k$, let $\textbf{S}_{\mathcal{L}_k}$ be the base interpolation function restricted to $\mathcal{L}_k$. Thus, the $k^{th}$ weak voter function for $x^*$ is obtained by applying the base interpolator $\textbf{S}_{(\mathcal{P}, \mathcal{M})}$ to estimate the value at $x^*$ using only the points in $\mathcal{L}_k$. \smallskip

Formally, the weak voter function set is defined as:
\begin{equation}\label{eq:Lk}
   f_k(x^*) = \textbf{S}_{\mathcal{L}_k}(x^*), \quad k = 1, \cdots, m
\end{equation} \smallskip

\subsection{Interpolation using the trained model}

Let us denote $x_k^* = f_k(x^*) = \textbf{S}_{\mathcal{L}_k}(x^*)$ as the $k^{th}$ weak voter function for $x^*$. Then, $\{x_k^*\}_{k=1}^{m}$ corresponds to the set of weak voter functions for $x^*$ resulting from all $m$ trees.

Thus, the interpolation function $\mathcal{Z}_{(\mathcal{P}, \mathcal{M})}$ corresponds to the aggregation of these weak voter functions:
\begin{equation}\label{eq:interpolator}
   e^* = \mathcal{Z}_{(\mathcal{P}, \mathcal{M})}(x^*) = G(\{x_k^*\}_{k=1}^{m})
\end{equation} \smallskip

The simplest choice for the aggregation function $G$ is the mean $\mathbb{E}[\cdot]$. In this case, the interpolation function becomes:
\begin{equation}\label{eq:interpolator_E}
    e_{\mathbb{E}}^* = \dfrac{1}{m}\sum_{k=1}^m x_k^*
\end{equation}  \smallskip

\subsection{Interpolation precision modelling}
\label{prec_model}

A precision model $p^*$ for $\mathcal{Z}_{(\mathcal{P}, \mathcal{M})}(x^*)$ can be defined using a loss function $\mathbb{L}$ as follows:
\begin{equation}\label{eq:precision}
p^* = \mathbb{E}_{\hat{P}(\{x_k^*\}_m)}(\mathbb{L}(e^*,\{x_k^*\}_m))
\end{equation} 

Equation \ref{eq:precision}represents a generalisation of the mean-variance concept within the context of Bayesian variability. \\

When using the mean-based interpolator $e_{\mathbb{E}}^*$, we can define an associated interpolation variance (or error) $\mathbb{V}_{e_{\mathbb{E}}^*}(x^*)$ as:
\begin{equation}\label{eq:interpolator_error}
   p_{\mathbb{E}}^* = \mathbb{V}_{e_{\mathbb{E}}^*}(x^*) = \dfrac{1}{m}\sum_{k=1}^m (x_k^* - e_{\mathbb{E}}^*)^2
\end{equation} \smallskip

\subsection{Rule of thumb for parameter choice}
\label{rule_thumb}

\subsubsection{When using Mondrian Forests}
The domain parameter $\Theta$ can be considered as any bounding box containing the positions of the conditioning data $\mathcal{P}$. Thus, in practice, the only two parameters of the model are: 
\begin{itemize}
\item The number of partitions (or tree structures) $m$. A reasonable suggestion for this parameter is that higher is better,  keeping in mind that higher values will directly impact time performance. Experiments have shown that certain stability is reached for $m \geq 500$ \citep{Egana2021ESI}, so this would be a good starting point.
\item The process lifetime $\lambda$. The only restriction for this parameter is that it must be positive. This renders the selection, or any sensitivity analysis, of its value challenging. In order to address this issue, a function of a normalised parameter $\alpha \in [0,1)$ is used to obtain suitable values for $\lambda$, which is defined as follows:
\begin{equation}\label{eq:lambda}
    \lambda(\alpha) = \dfrac{1}{\mu(\Theta)(1 - \alpha)}
\end{equation}

In practice, $\alpha$ controls the average tree depth in the forest, determining how finely the space is partitioned. In this way, $\alpha = 0$ will generate the coarsest partition, while $\alpha \rightarrow 1$ will generate finer ones. 
$\alpha$ must be carefully chosen to ensure that the base interpolation function $\textbf{S}_{\mathcal{L}_k}$ has sufficient sample data. Once $m$ has been defined, it is recommended to use cross-validation to find the optimal $\alpha$, typically within $[0.7, 0.95]$.
\end{itemize}

\subsubsection{When using Voronoi Forests}

As seen in Mondrian Forest, the domain parameter $\Theta$ may be regarded as any bounding box encompassing $\mathcal{P}$. Consequently, the parameters of the model remain identical, yet their respective roles and practical considerations differ:

\begin{itemize}
\item The number of partitions (or tree structures) $m$. A reasonable suggestion for this parameter is that higher is better, keeping in mind that higher values will directly impact time performance. However, the possibility of reaching stability for a certain value of $m$ is yet to be studied in the context of Voronoi Forests.

\item The process lifetime $\lambda$. This parameter determines the expected number of Voronoi nuclei for each partition. 
In the context of Voronoi Forest, $\lambda$ is related to the mean of the Poisson distribution. Although the only theoretical restriction for this parameter is that it must be greater than one, in Spatialize, it is also constrained to match the expected number of leaves in a Mondrian tree. To this end, $\lambda$ is calculated by multiplying the parameter $\alpha$, defined in the context of Mondrian forests, by a factor according to the number of observations $N_s$, as follows:
\begin{equation}\label{eq:lambda_vor}
    \lambda(\alpha) = \frac{1}{2} *  N_s * \alpha, \alpha \in [0,1)
\end{equation}

As in the case of Mondrian partitions, $\alpha$ also controls how coarse or fine the partitions are by affecting the number of Voronoi nuclei. Thus, $\alpha = 0.25$ will generate the coarsest partition, while $\alpha \rightarrow 1$ will generate finer ones. 
\end{itemize}

In the context of sampling from a trained tree $\text{VT} (\Theta, \lambda| \mathcal{P})$, additional considerations arise with respect to $m$ and $\lambda$. 
Firstly, the number of nuclei, $K(\lambda(\alpha))$, must not exceed the number of observations, $N_s$. 
Secondly, as $K(\lambda(\alpha)) \rightarrow N_s$, when sampling from $\mathcal{P}$, the number of possible Voronoi nuclei combinations. Therefore, a large $m$ value may result in duplicated partitions. To avoid such inefficiencies, it is recommended that $m \ll C(N_s, K)$, or alternatively, not employing trained trees (also possible on Spatialize). Alternatively, Spatialize also permits the direct sampling of $\Phi_K$ from $\Theta$
(i.e. not employing trained trees).

\section{Usage examples}
\label{use_ex}

The main motivation behind the development of the \verb|spatialize| library was to provide the scientific and technical community with a robust and automated spatial estimation tool that can be used across different disciplines by researchers and professionals who are not experts in geostatistics. 

In this section, we present a sequence of usage examples which illustrate how \verb|spatialize| facilitates the task of automated spatial estimation, including hyper-parameter search functions and estimation functions, as described in detail in the Appendix \ref{App:user-manual}.

Furthermore, these examples demonstrate that the efficacy of our tool is comparable to or superior to that of other automatic spatial estimation tools, taking into account parameters obtained through automated grid searches -- a process that does not necessitate the involvement of an expert selection a priori. To make the point, examples of estimation generated with the \verb|SciPy| library are also presented.

\subsection{Gridded data estimation} \label{grid_est}

Estimation with ESI on data that is on a regular grid is performed with the \verb|esi_griddata()| function. Two examples have been developed and are described below. 

Firstly, we present a use case in which the reference two-dimensional surface is a cubic-type function. The idea of this example is to compare \verb|spatialize| with various types of spatial estimation within the \verb|SciPy| library, using a discrete sampling of a cubic-type function, while also illustrating the estimation process with gridded data using the \verb|spatialize| library. Then, a comparative analysis of the two partitioning methods (Mondrian and Voronoi) is presented. Furthermore, we introduce the \verb|esi_hparams_search| function, which searches for optimal ESI parameters, and the \verb|loss| functionality, which allows for the definition of custom loss functions.

\begin{figure}[H]
    \centering
    \includegraphics[width=0.4\linewidth]{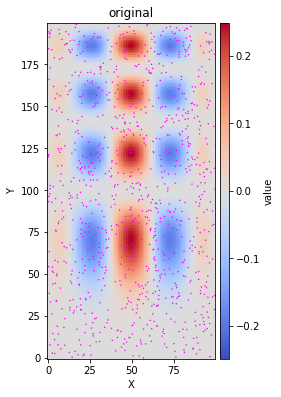}
    \caption{The original cubic type function, plotted in superposition to the randomly generated points from which the spatial interpolations will be made.}
    \label{fig:original_plus_poins}
\end{figure}

The Code snippet \ref{listing:1} shows how to generate the grid and some points and the corresponding values generated by the cubic type function -- taken from the documentation of the \verb|griddata| function of the module \verb|scipy.interpolate|, included in the \verb|SciPy| library. This dataset is used as input for all examples performing estimations where data are included in a regular grid. Figure \ref{fig:original_plus_poins} illustrates this function alongside randomly sampled points, which are to be used in the comparison of the different interpolation methods.

\begin{listing}[H]
\begin{footnotesize}
\begin{verbatim}
def func(x, y):  # a kind of "cubic" function
return x * (1 - x) * np.cos(4 * np.pi * x) * 
    np.sin(4 * np.pi * y ** 2) ** 2

grid_x, grid_y = np.mgrid[0:1:100j, 0:1:200j]

rng = np.random.default_rng()
points = rng.random((1000, 2))
values = func(points[:,0], points[:,1])   
\end{verbatim}
\end{footnotesize}
\caption{Generation of the points and values that are input to the spatial gridded estimation examples presented below.}
\label{listing:1}
\end{listing}

\subsubsection{ESI vs scipy}
\begin{figure}[H]
    \centering
    \includegraphics[width=1\linewidth]{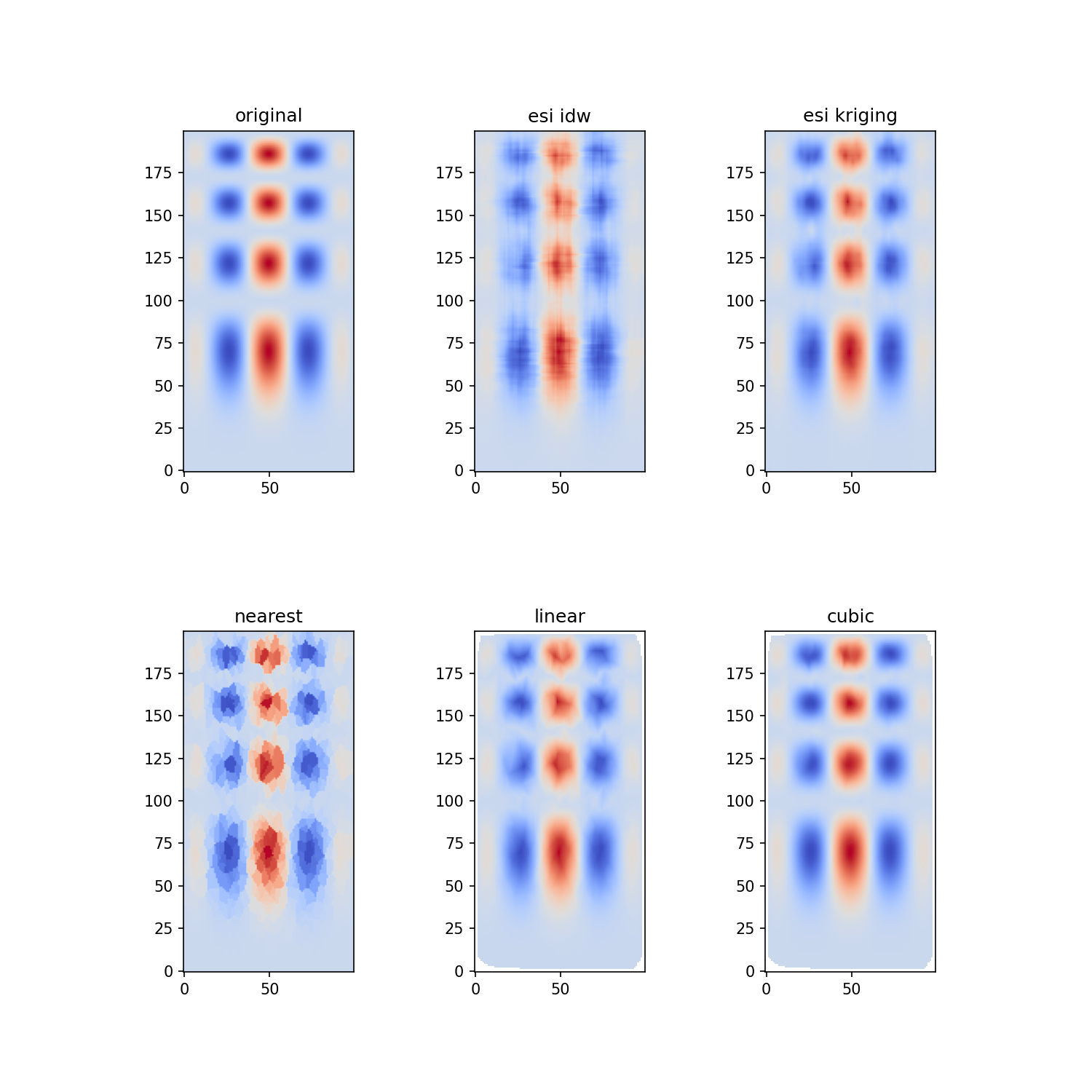}
    \caption{Comparison of ESI and SciPy estimates for the same gridded data set}
    \label{fig:esi_scipy_comp}
\end{figure}

As mentioned above, in order to have a comparison with the results of known interpolation tools implemented in Python, we use the \verb|griddata| function of the \verb|scipy.interpolate| module with the ‘nearest neighbour’, ‘linear’ and ‘cubic’ methods. Code snippet \ref{listing:2} shows how to estimate with this function the three cases that are shown in Figure \ref{fig:esi_scipy_comp}.

\begin{listing}[H]
\begin{footnotesize}
\begin{verbatim}
from scipy.interpolate import griddata
nearest_result=griddata(points, values, (grid_x, grid_y), method='nearest')
linear_result=griddata(points, values, (grid_x, grid_y), method='linear')
cubic_result=griddata(points, values, (grid_x, grid_y), method='cubic')
\end{verbatim}
\end{footnotesize}
\caption{Griddata function to generate three different estimations using SciPy library.}
\label{listing:2}
\end{listing}

Then, we use the \verb|esi_griddata| estimation function with the local IDW and Kriging interpolators and arbitrary non-optimal parameter sets. Code snippets \ref{listing:3} and \ref{listing:4} show the two implementations. 

\begin{listing}[H]
\begin{footnotesize}
\begin{verbatim}
esi_griddata(points, values, (grid_x, grid_y),
             local_interpolator="idw",
             p_process="mondrian",
             data_cond=False,
             exponent=1.0,
             n_partitions=500, alpha=0.985,
             agg_function=af.mean)
\end{verbatim}
\end{footnotesize}
\caption{Gridded estimation using IDW as local interpolator.}
\label{listing:3}
\end{listing}

\begin{listing}[H]
\begin{footnotesize}
\begin{verbatim}
esi_griddata(points, values, (grid_x, grid_y),
             local_interpolator="kriging",
             model="spherical", nugget=0.0, range=10.0,
             n_partitions=500, alpha=0.9,
             agg_function=af.mean)  
\end{verbatim}
\end{footnotesize}
\caption{Gridded estimation using Kriging as local interpolator.}
\label{listing:4}
\end{listing}

In Figure \ref{fig:esi_scipy_comp}, we can see the original shape generated with the cubic function and the estimations made with ESI, based on IDW and Kriging as local interpolators and the three interpolation options generated with the SciPy library. It can be observed that the results produced by ESI, without parameter optimisation and without any structural assumption, are quite acceptable compared to those generated by the SciPy interpolators, which have a structural assumption defined a priori. Obviously, if one introduces an inductive bias by somehow knowing that the underlying function to be interpolated is cubic, the cubic model will have the best result, at least visually. We can call this the inductive bias effect.

As mentioned within Section \ref{weak_voter}, both Mondrian Forest and Voronoi Forest can be used as the partitioning method in the case of ESI-IDW\footnote{For the case of the local kriging interpolator, \texttt{spatialize} only has the Mondrian Forest implementation.}. Code snippet \ref{listing:5} shows how to generate both versions of estimation, in the gridded case, with the IDW interpolator.

\begin{listing}[H]
\begin{footnotesize}
\begin{verbatim}
esi_griddata(points, values, (grid_x, grid_y),
             local_interpolator="idw",
             p_process="mondrian",
             data_cond=False,
             exponent=1.0,
             n_partitions=500, alpha=0.985,
             agg_function=af.mean
             )

esi_griddata(points, values, (grid_x, grid_y),
             local_interpolator="idw",
             p_process="voronoi",
             data_cond=True, # and False alternatively
             exponent=1.0,
             n_partitions=500, alpha=0.985,
             agg_function=af.mean
             )
\end{verbatim}
\end{footnotesize}
\caption{Gridded estimation using IDW as local interpolator and using both Mondrian and Voronoi partition methods.}
\label{listing:5}
\end{listing}

\begin{figure}[H]
    \centering
    \includegraphics[width=1\linewidth]{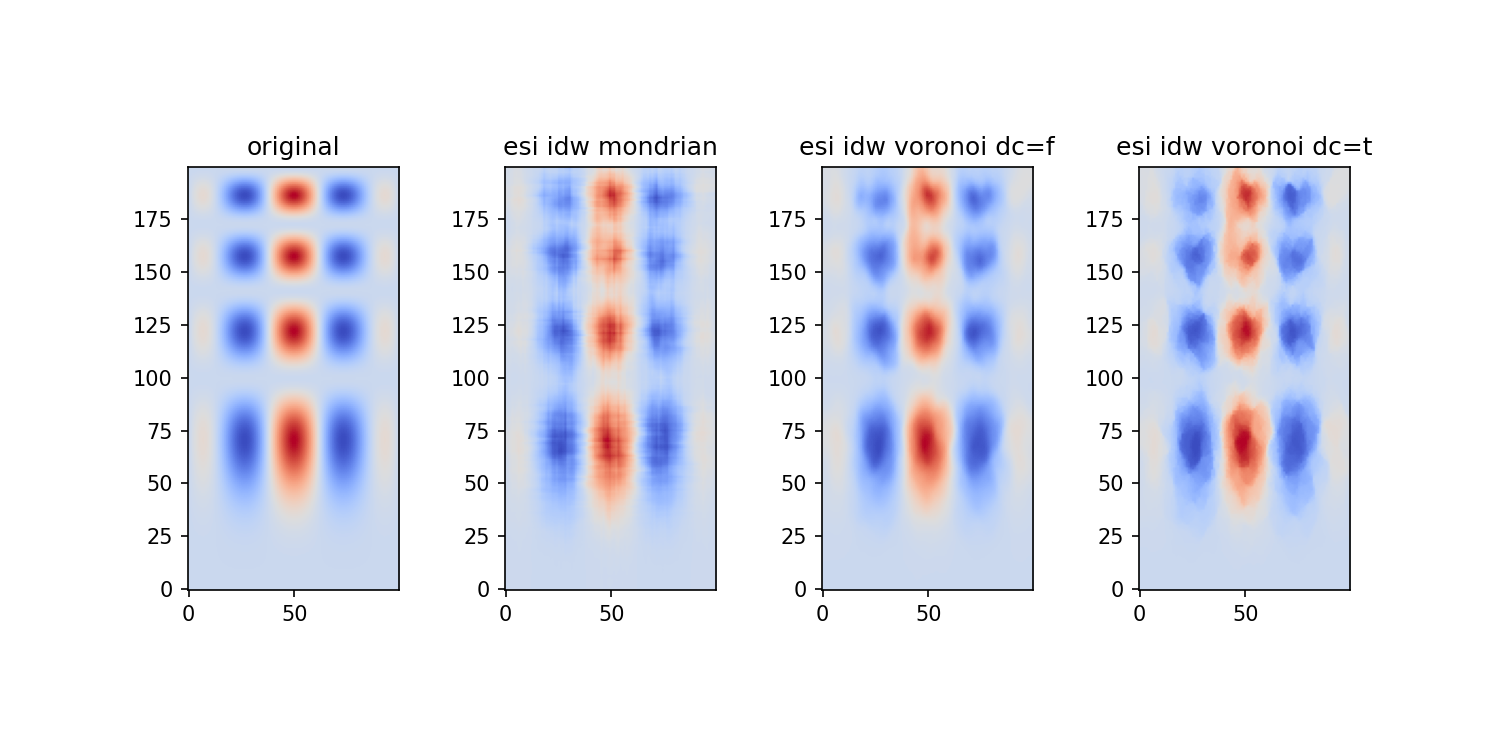}
    \caption{Comparison of ESI-IDW estimates for Mondrian and Voronoi partition methods, for the two graphs on the right $dc = data\_cond$ represents whether there is (t) or not (f) conditioning on the data.}
    \label{fig:idw_mond_vor}
\end{figure}

Figure \ref{fig:idw_mond_vor} shows both ESI-IDW partitioning alternatives: Mondrian Forest and Voronoi Forest. In the latter case, the case without and with data conditioning is shown. It can be seen that activating data conditioning makes the estimation look more like the case where Mondrian Forest is used for partitioning, while not using data conditioning makes the estimation look smoother, i.e., closer to the original scenario. This situation looks interesting because it reflects the fact that the partitioning process can significantly affect the outcome. Surely, there must be some Bayesian generative argument behind this, which clearly deserves more attention.

\subsubsection{Hyper parameter search}
\label{grid_est_param}

One of the powerful elements of the \verb|spatialize| library is the facility to automate the search for the best parameters for ESI estimation. This search is performed by the function \verb|esi_hparams_search|, which is available for both the gridded and ungridded estimation cases. 
This function employs cross-validation to determine the parameter combination that yields the minimum error out of a previously defined set.

The purpose of this section is to provide an example of the use of the grid search functions, focusing on the gridded case. In this example, we first conduct a search from a set of options for each of the parameters, and then perform the estimation with the best parameters found. We then compare the results with the best estimate that can be achieved using the IDW estimator as a global estimator without ESI, to give a reference for the improvement implied by using it only as a local interpolator in ESI. In the case of IDW without ESI, the \verb|spatialize| library includes a grid search function, which is analogous to the one for ESI estimation parameters search.

A parameter search is now performed on the same two-dimensional data set that was previously used, employing the \verb|esi_hparams_search| function. As can be seen in Code snippet \ref{listing:6}, the search function receives ranges or sets of arguments, which specify different combinations of parameters for interpolation. In the example, Kriging is used as the local interpolator, with four different omnidirectional variogram model options. The function generates and compares all possible scenarios with the combinations of these parameters.

\begin{listing}[H]
\begin{footnotesize}
\begin{verbatim}
search_result = 
esi_hparams_search(points, values, (grid_x, grid_y),
                   local_interpolator="kriging", griddata=True, k=10,
                   model=["spherical", "exponential", "cubic", "gaussian"],
                   nugget=[0.0, 0.5, 1.0],
                   range=[10.0, 50.0, 100.0, 200.0],
                   alpha=[0.97, 0.96, 0.95])
\end{verbatim}
\end{footnotesize}
\caption{Grid search to find the best case parameters within the given ranges and options for optional named arguments in esi\_hparams\_search function, in this case for Kriging local interpolator.}
\label{listing:6}
\end{listing}

\begin{figure}[H]
    \centering
    \includegraphics[width=1\linewidth]{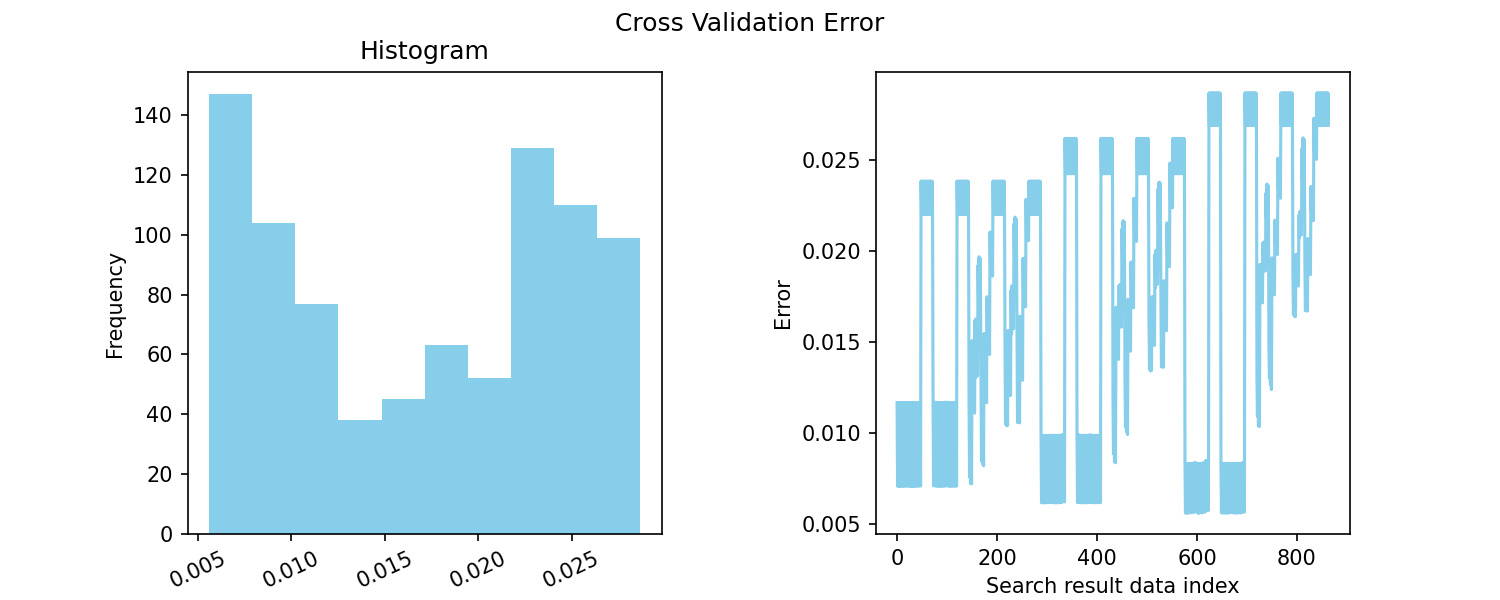}
    \caption{Cross validation error for grid search using Kriging as local interpolator}
    \label{fig:cross_val_grid_krig}
\end{figure}

Figure \ref{fig:cross_val_grid_krig} shows the error frequency graphs (left) and the different error levels for each scenario (right). In Code snippet \ref{listing:6}, the object \verb|search_result| will contain the result data for the grid search executed. 
It is interesting to observe the result in this graphical way, where one can see how the level of estimation error is distributed (histogram on the left) and then how this error evolves in the sequence of scenarios running during the search. 

Finally, the estimation is performed based on the parameters of the scenario with the lowest cross-validation error, which in this case has an index of 302. Since this is an interpolation, and therefore, the estimate at points where there is grid data is by construction equal to the reference data, to calculate the error, $K$ points are removed in a K-Fold round and the error of the estimate at those points is calculated, and then the average of all iterations is obtained. In this example, $K=10$. It can be noted in the code below that the gridded estimation function only requires as optional arguments those provided by the method \verb|search_result.best_result()|.

\begin{listing}[H]
\begin{footnotesize}
\begin{verbatim}
result = esi_griddata(points, values, (grid_x, grid_y),
                      best_params_found=search_result.best_result()
                      )
\end{verbatim}
\end{footnotesize}
\caption{Gridded estimation using best result of esi\_hparams\_search result.}
\label{listing:7}
\end{listing}

\begin{figure}[H]
    \centering
    \includegraphics[width=0.7\linewidth]{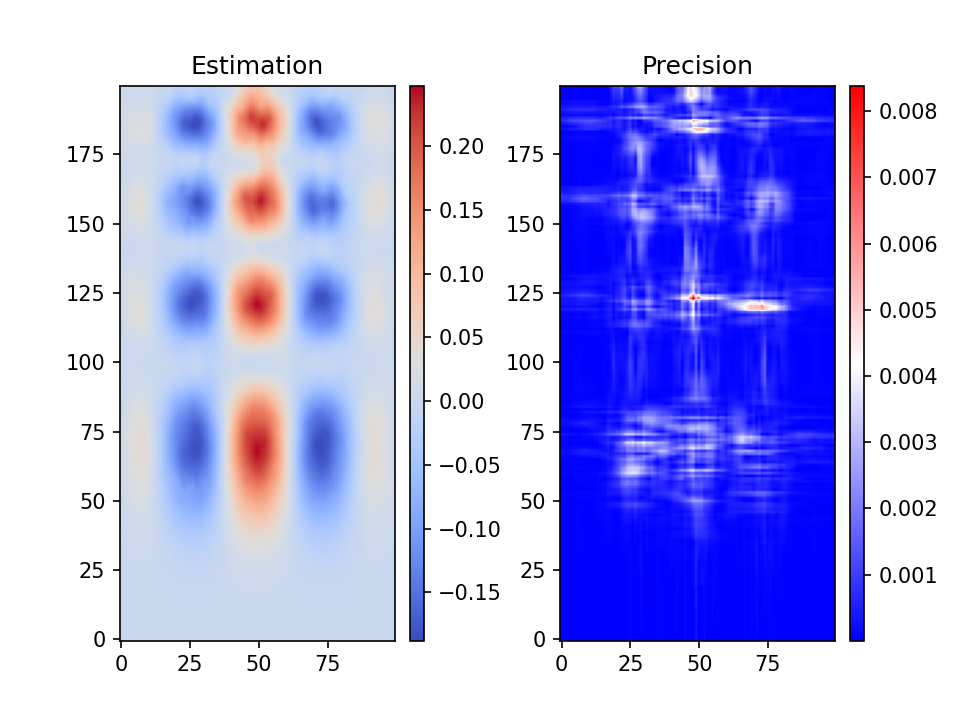}
    \caption{ESI grid search best estimation using Kriging as local interpolator (left) and precision obtained with MSE (default) loss function (right)}
    \label{fig:grid_search_grid_krig}
\end{figure}

Figure \ref{fig:grid_search_grid_krig} shows the result of the estimation with the best parameters obtained from the previous search (left). The model that delivered the search is the ‘spherical’ one, with an $alpha=0.95$. It is remarkable how the structure of the original image is recovered, considering that there are no domains or variographic studies involved in this model. On the right is the accuracy, calculated with the default loss function (MSE). It can be seen that the largest impressions appear around the boundaries of the structures.

Now, to show how powerful this spatial interpolation method is, we compare the best-case estimations of ESI with IDW as the base interpolator and an interpolation that only uses IDW. The aim is to show that this method does have the ability to rescue the structural aspects of the presented example, provided that they are regular shapes (cubic type function, in this case).

First, we search for optimal IDW hyperparameters using the \verb|idw_hparams_search| function, analogous to the \verb|esi_hparams_search| function. Code snippet \ref{listing:8} shows its implementation.

\begin{listing}[H]
\begin{footnotesize}
\begin{verbatim}
search_result = 
idw_hparams_search(points, values, (grid_x, grid_y),
                   griddata=True, k=10,
                   radius=[0.07, 0.08],
                   exponent=(0.001, 0.01, 0.1, 1, 2)
                   )
\end{verbatim}
\end{footnotesize}
\caption{Grid search for IDW estimation (without ESI).}
\label{listing:8}
\end{listing}

\begin{figure}[H]
    \centering
    \includegraphics[width=1\linewidth]{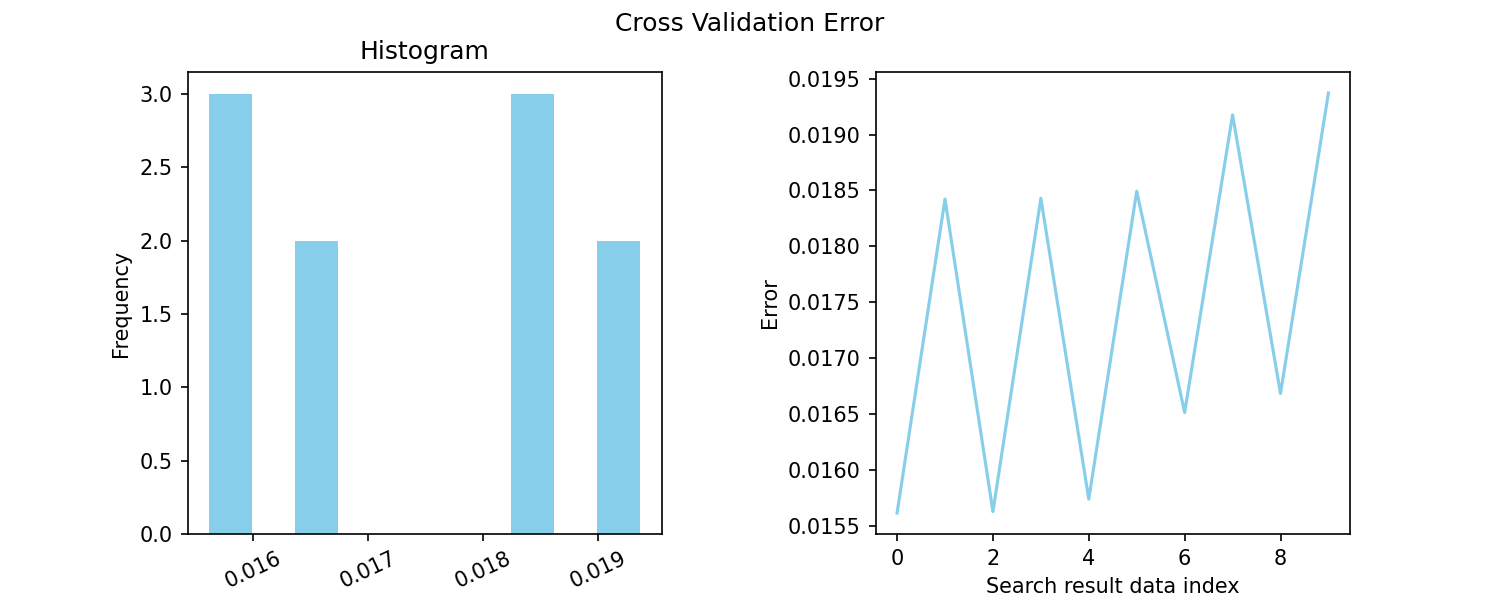}
    \caption{Cross validation error for grid search for IDW interpolator without ESI.}
    \label{fig:basic_idw_grid_search}
\end{figure}

As shown in Figure \ref{fig:basic_idw_grid_search}, the minimum errors obtained are in the order of $0.016$.

We can then generate the best estimate from the best hyperparameters found in the search, as shown in code snippet \ref{listing:9}.

\begin{listing}[H]
\begin{footnotesize}
\begin{verbatim}
result = 
idw_griddata(points, values, (grid_x, grid_y),
             best_params_found= search_result.best_result(optimize_data_usage=False))
\end{verbatim}
\end{footnotesize}
\caption{Gridded estimation with IDW without ESI, using IDW grid search best result.}
\label{listing:9}
\end{listing}

\begin{figure}[H]
    \centering
    \includegraphics[width=0.9\linewidth]{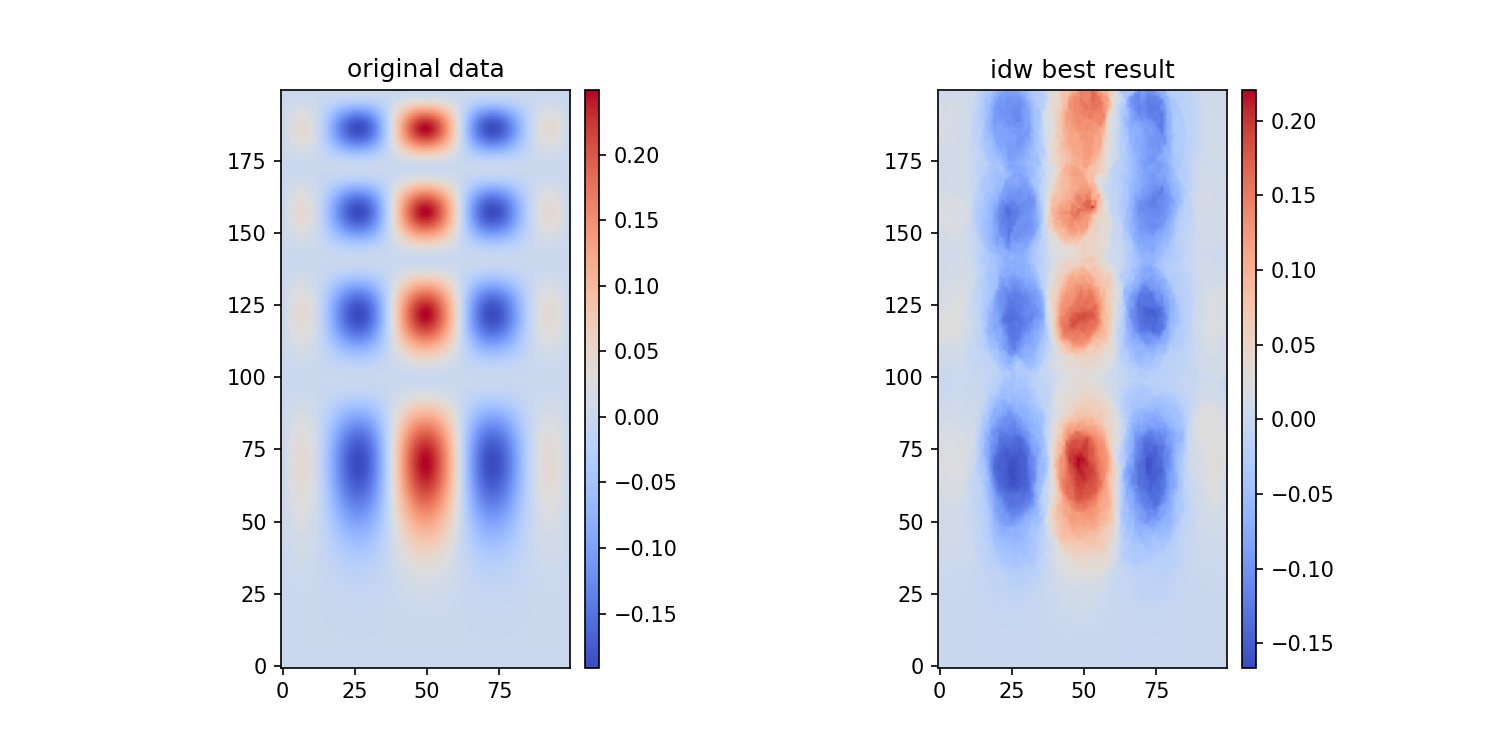}
    \caption{Best estimation using IDW interpolator without ESI}
    \label{fig:basic_idw_estimation}
\end{figure}

Figure \ref{fig:basic_idw_estimation} shows the best possible estimation for the search parameter grid performed in the pure IDW case. It can be seen that the IDW interpolator does rescue the original image structures.

Now, we will use IDW as a local interpolator for ESI, performing the same parameter search and estimation exercise. The code shown in Code snippet \ref{listing:10} allows us to do this operation. For the parameter search, we have defined a wide set of combinations, where we even consider different aggregation functions to be applied on the sets of ESI scenes obtained in each case. In addition, the Voronoi partitioning method has been used since it generates more regular partition elements, which may be favourable in this case, given that the local IDW interpolator works radially.

\begin{listing}[H]
\begin{footnotesize}
\begin{verbatim}
search_result = 
esi_hparams_search(points, values, (grid_x, grid_y),
                   local_interpolator="idw", griddata=True, k=10,
                   p_process="voronoi",
                   n_partitions=(30, 50, 100),
                   exponent=[0.001, 0.01, 0.1, 1, 2],
                   alpha=(0.95, 0.97, 0.98, 0.985),
                   agg_function={"mean": af.mean,
                                 "median": af.median,
                                 "p25": af.Percentile(25),
                                 "p75": af.Percentile(75)
                                 })
\end{verbatim}
\end{footnotesize}
\caption{Grid search to find the best case parameters within the given ranges and options for optional named arguments in esi\_hparams\_search function, in this case for IDW local interpolator with gridded data ESI estimation.}
\label{listing:10}
\end{listing}

\begin{figure}[H]
    \centering
    \includegraphics[width=1\linewidth]{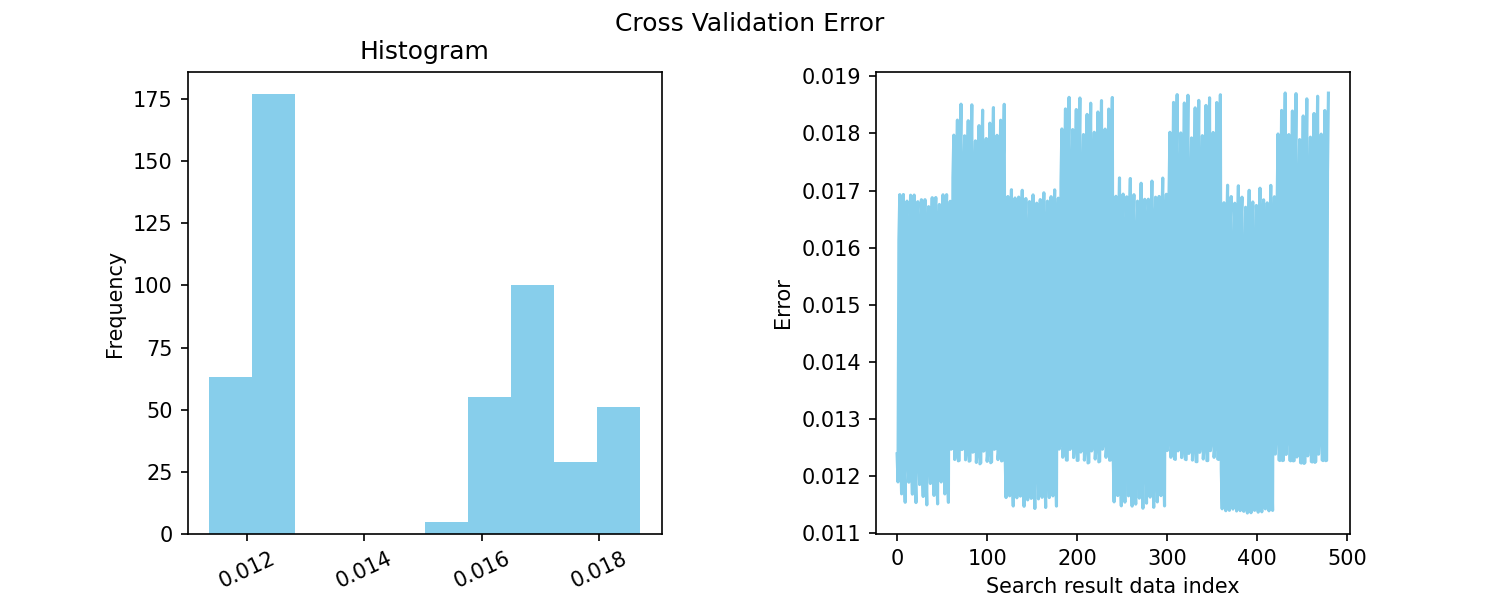}
    \caption{Cross validation error for grid search using IDW as local interpolator}
    \label{fig:cross_val_grid_idw}
\end{figure}

Figure \ref{fig:cross_val_grid_idw} shows the errors for the 400+ scenarios generated by the search. The minimum errors obtained are in the order of $0.012$. In this case, the power of ESI is expressed in 25\% lower error levels than in the pure IDW case (Figure \ref{fig:basic_idw_grid_search}).

Finally, analogous to the case of ESI-Kriging estimation, code snippet \ref{listing:11} performs the ESI-IDW estimation with the parameters found in the search.

\begin{listing}[H]
\begin{footnotesize}
\begin{verbatim}
result = esi_griddata(points, values, (grid_x, grid_y),
                      best_params_found=search_result.best_result()
                      )
\end{verbatim}
\end{footnotesize}
\caption{Gridded estimation with IDW using grid search best result.}
\label{listing:11}
\end{listing}

\begin{figure}[H]
    \centering
    \includegraphics[width=0.9\linewidth]{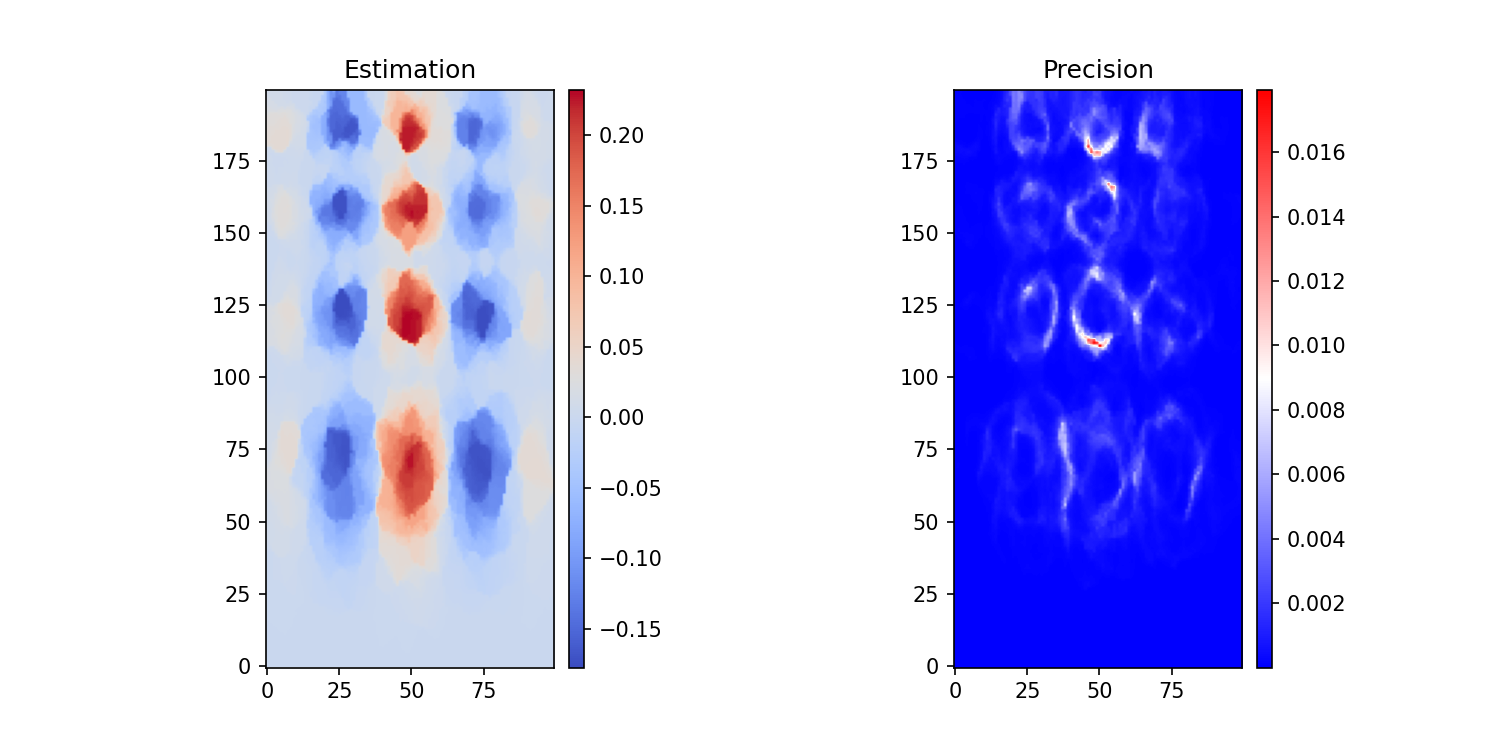}
    \caption{ESI grid search best estimation using IDW as local interpolator (left) and precision obtained with Operational Error function (right)}
    \label{fig:grid_search_grid_idw}
\end{figure}

Finally, Figure \ref{fig:grid_search_grid_idw} presents the estimation with the best parameters found by the grid search above. Although the accuracy in this case is a little lower than in the case of the ESI-Kriging estimation, this result is noticeably better than the one presented with the basic IDW interpolation. It can be seen that capturing the structures of the original function is a result of the power of ESI.

\subsubsection{Custom precision functions}
\label{custom_p}

In this section, we will review the process of implementing accuracy metrics through relationships other than mean-variance.

If we would like to implement our own loss function for the calculation of custom precision from the scenarios generated by the ESI estimator and a particular aggregate estimate, \verb|spatialize| provides a very modular and convenient way to do so. For example, to have our own implementation of the operational error, which is also implemented as a class in the library (module \verb|lossfunction|), we would use the following code contained in the example mentioned above.

\begin{listing}[H]
\begin{footnotesize}
\begin{verbatim}
from spatialize.gs.esi.lossfunction import loss

def op_error_precision(estimation, esi_samples):
    dyn_range = np.abs(np.nanmin(esi_samples) - np.nanmax(esi_samples))

    @loss(af.mean)
    def _op_error(x, y):
        return np.abs(x - y) / dyn_range

    return _op_error(estimation, esi_samples)
\end{verbatim}
\end{footnotesize}
\caption{Example of how to create a customised loss function, in this case, a version of the operational error.}
\label{listing:21}
\end{listing}

\begin{figure}[H]
    \centering
    \includegraphics[width=1\linewidth]{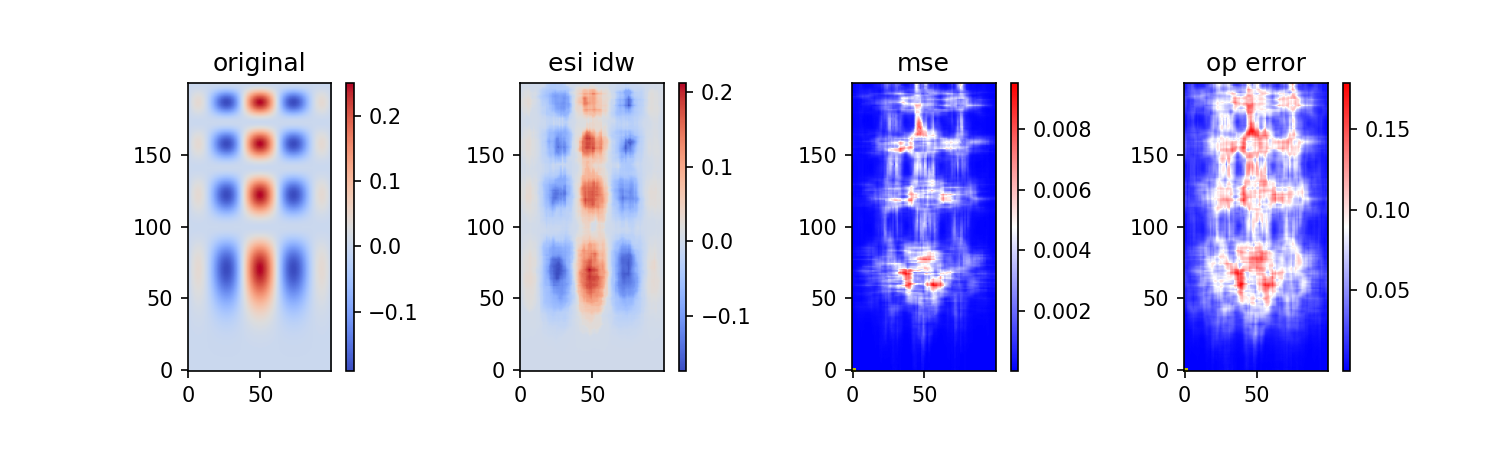}
    \caption{Mean square error and operational error for an ESI-IDW interpolation}
    \label{fig:cust_prec_idw}
\end{figure}

\begin{figure}[H]
    \centering
    \includegraphics[width=1\linewidth]{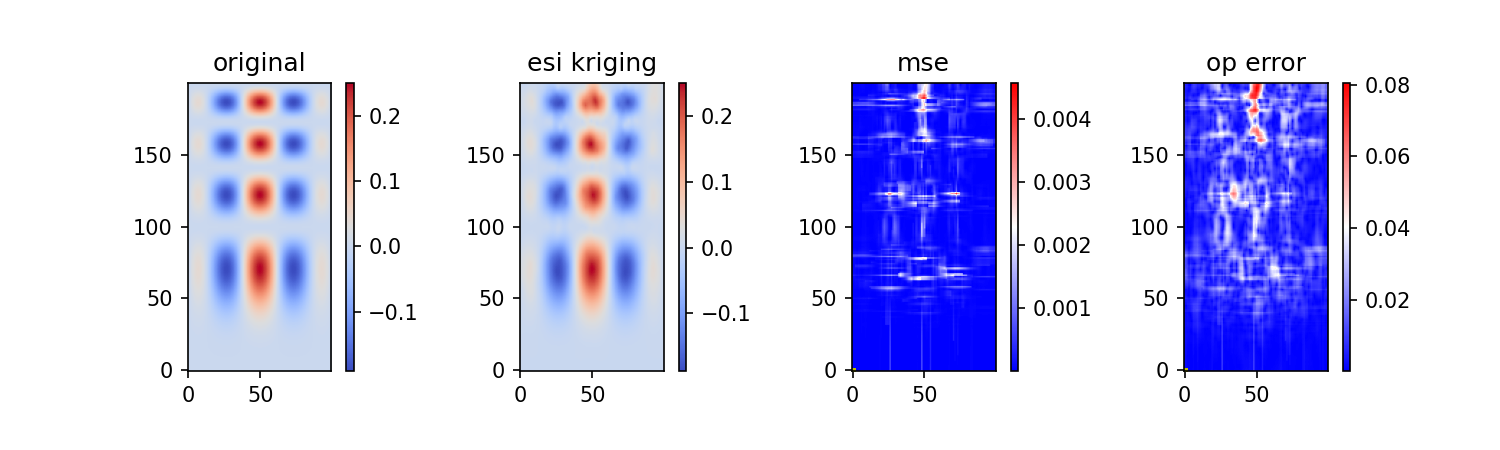}
    \caption{Mean square error and operational error for an ESI-Kriging interpolation}
    \label{fig:cust_prec_krig}
\end{figure}

As can be seen in the above code, the function is defined, and within it, a second function is defined, which is decorated by \verb|@loss()|, within which is also assigned the aggregation function that will be used to aggregate the unit loss calculations per scenario into a single precision layer. Figures \ref{fig:cust_prec_idw} and \ref{fig:cust_prec_krig} show a comparison between the default accuracy contained in the \verb|ESIResult| object and that produced by our operational error function for the ESI-IDW (in this case, using Mondrian partitions) and ESI-Kriging cases, respectively. 

\subsection{Non-gridded data estimation}\label{non-grid_est}

A fundamental advantage of the ESI model in comparison to traditional geostatistics is its capacity to effortlessly analyse d-dimensional\footnote{Currently, spatialize offers support for up to two dimensions when using Voronoi partitions and five dimensions when using Mondrian. For the 4D (space-time) and 5D (spatial with two angles, for fault description, for example) case, the implementation includes only IDW as local ESI interpolator.}, non-gridded data. 
As a result, the \verb|spatialize| library is able to automatically generate spatial estimates for any set of points in space, even when these are not arranged on a regular grid. This includes irregularly-spaced points, incomplete grids, and 2D surfaces with variations on a third axis (frequently termed 2.5D). In this sense, it is highly flexible. 

In this section, we employ the \verb|esi_nongriddata()| function, which generates estimates from a set of sample points at a set of unmeasured points at arbitrary locations in space. 
Employing non-gridded sample data\footnote{In this case, the estimates will be calculated on a set of points arranged as a grid so that a comparison with Ordinary Kriging is possible. However, Spatialize could be implemented over any given set of points.}, we will present a comparison of the estimates derived from three different Kriging implementations and ESI-Kriging (i.e. using Universal Ordinary Kriging as base interpolator).

The dataset employed for this example, as well as the corresponding Ordinary Kriging estimates, are available to be loaded in the \verb|spatialize| library, as shown in Code snippet \ref{listing:12}.

\begin{listing}[htb]
    \begin{footnotesize}
    \begin{verbatim}
        from spatialize.data import load_drill_holes_andes_2D
        
        samples, locations, krig, _ = load_drill_holes_andes_2D()
    \end{verbatim}
\end{footnotesize}
\caption{Loading the samples, locations, and ordinary kriging estimate for the drill\_holes\_andes\_2D dataset in spatialize.}
\label{listing:12}
\end{listing}

This is a set of $400$ copper grade data, placed non-regularly, with coordinates in the space of $60,000$ points.

\subsubsection{ESI vs Kriging}
To produce the expert estimate using Ordinary Kriging in the example, an omnidirectional experimental variogram was generated, used and fitted to a function (theoretical variogram) containing two nested spherical structures (Equation \ref{eq:non-grid-variogram}).

\begin{equation} \label{eq:non-grid-variogram}
\gamma(h) = \text{sill}_1 \left( 1.5 \frac{h}{\text{range}_1} - 0.5 \left( \frac{h}{\text{range}_1} \right)^3 \right) + \text{sill}_2 \left( 1.5 \frac{h}{\text{range}_2} - 0.5 \left( \frac{h}{\text{range}_2} \right)^3 \right)    
\end{equation}

Where:
\[
\text{sill}_1 = 0.17, \quad \text{range}_1 = 95, \quad \text{sill}_2 = 0.14, \quad \text{range}_2 = 220
\]

\begin{figure}[htb]
    \centering
    \includegraphics[width=0.6\linewidth]{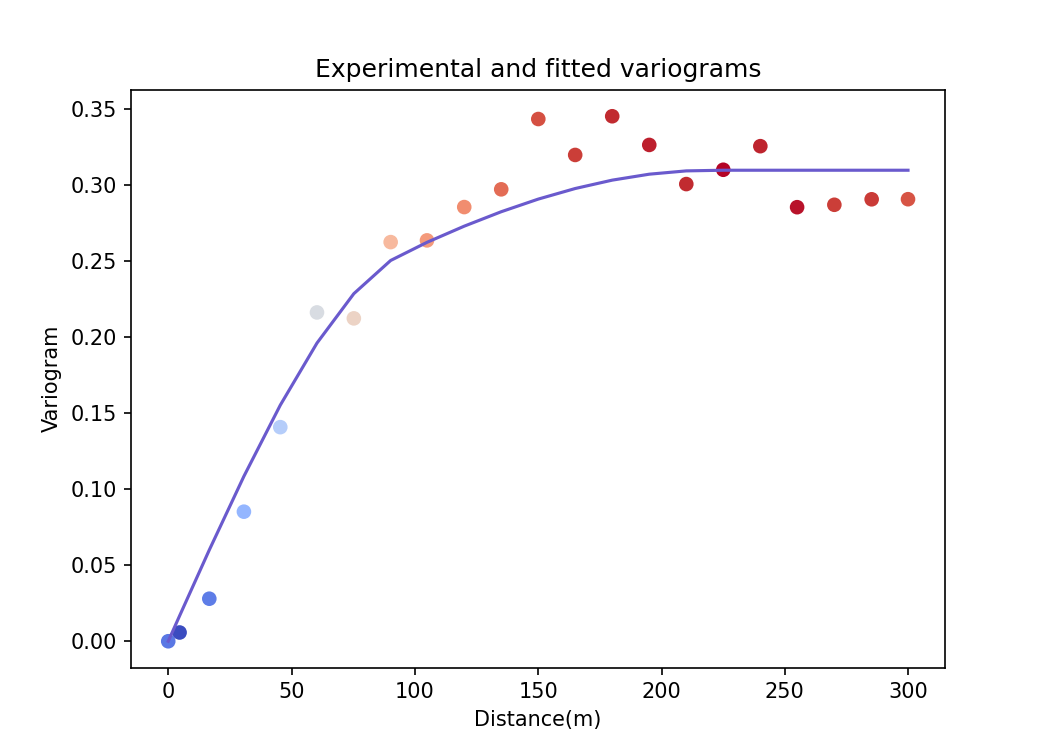}
    \caption{Experimental (dots) and theoretical (blue line) variograms for the Global Ordinary Kriging example.}
    \label{fig:variogram}
\end{figure}

Both experimental and theoretical variograms are shown in Figure \ref{fig:variogram}.

\begin{figure}[htb]
    \centering
    \includegraphics[width=1\linewidth]{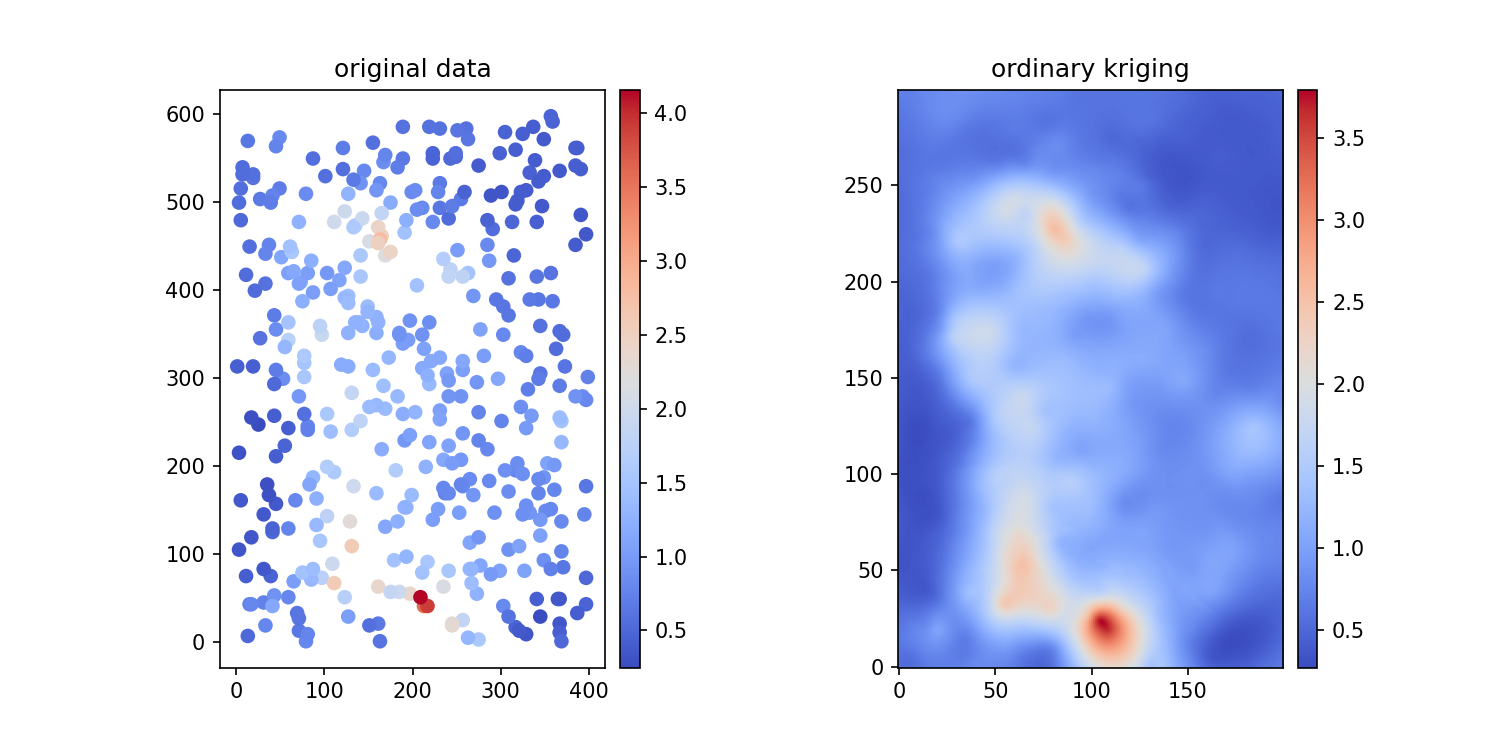}
    \caption{Map of points with values that are the input for the estimation in non-gridded data (left), and a Global Ordinary Kriging estimation based on those data points (right).}
    \label{fig:original_no_grid_data}
\end{figure}

 Figure \ref{fig:original_no_grid_data} (left) shows the data points (left), and the Global Ordinary Kriging estimation (right), which corresponds to the example developed in \citet{Egana2021ESI}.

\begin{figure}[htb]
    \centering
    \includegraphics[width=1\linewidth]{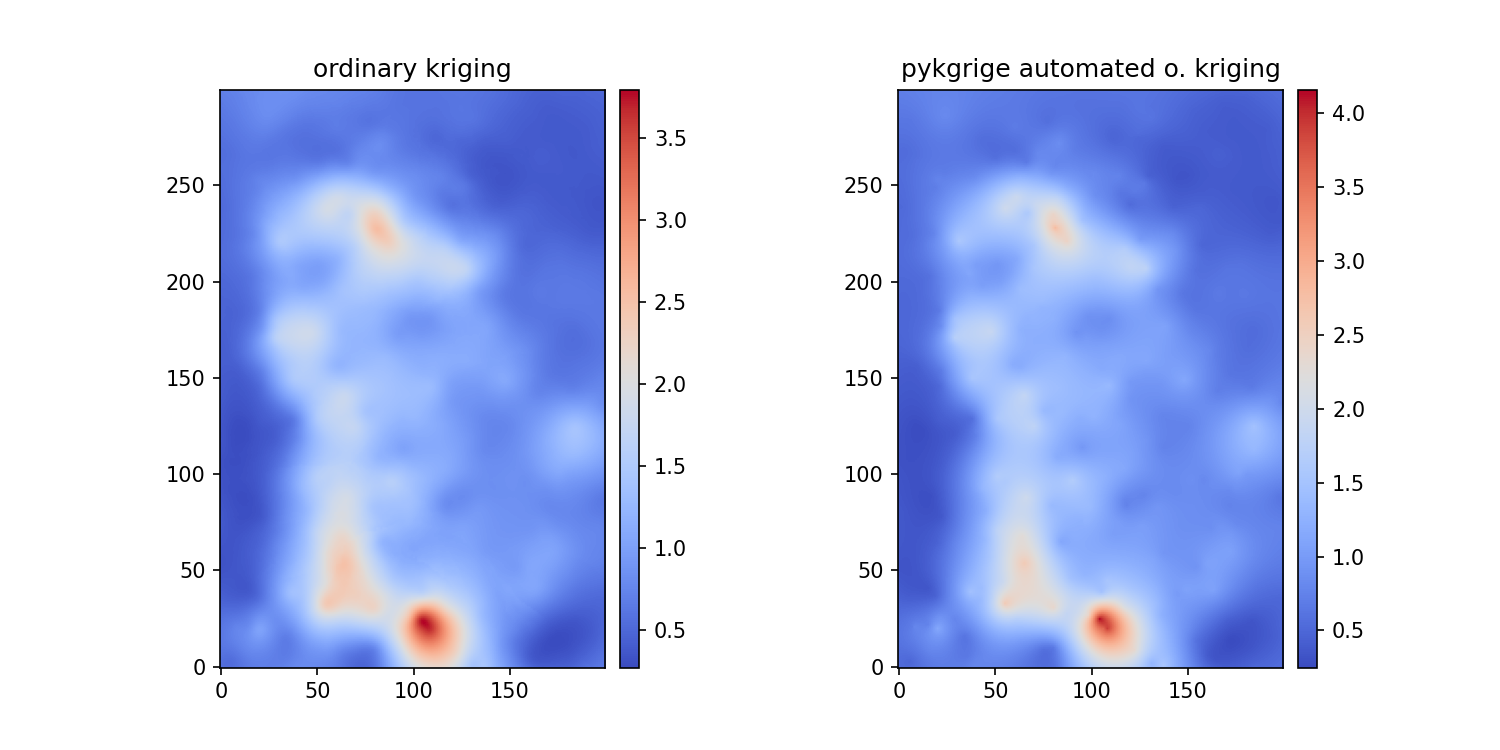}
    \caption{Global Ordinary Kriging estimation based on non-gridded example data points (left), and an automated Kriging estimation with \textit{pykrige} (right).}
    \label{fig:krig_and_pykrige}
\end{figure}

In addition, we have implemented an automated workflow that uses \verb|scikit-learn| to run a parameter grid search, in order to obtain the best variogram model and fit the experimental variogram, and \verb|PyKrige| to run Ordinary Kriging using the best parameters found. Figure \ref{fig:krig_and_pykrige} shows the results for the automated implementation (right) alongside the manual expert implementation of Global Ordinary Kriging (left).

This automated approach differs from the process carried out by the expert in three ways:
It uses global Kriging instead of domains depending on local stationarities; it determines variogram parameters in a heuristic way rather than through expert judgment; 
and it assumes that the variogram is the same in all directions (isotropy), ignoring potential anisotropies in the data.

In this example, the result obtained for the parameter search is as follows:
\begin{footnotesize}
    \begin{verbatim}
best_score R² = 0.918
best_params = {'variogram_model': 'exponential'}
\end{verbatim}
\end{footnotesize}

As we can see in Figure \ref{fig:krig_and_pykrige}, the estimation obtained with the automated algorithm is comparable to the ordinary Kriging performed manually on the basis of expert judgment. This is because the example is an ideal case of isotropy and stationarity, i.e. it fulfils the basic assumptions of a linear method such as kriging.

Next, the same estimation previously performed with manual and automated Kriging was performed using the \verb|spatialize| library.

First, the \verb|esi_hparams_search()| function was employed to obtain the best parameters for the estimation. The evaluated set is shown in the following snippet:

\begin{footnotesize}
\begin{verbatim}
search_result =  
esi_hparams_search(points, values, xi,
                   local_interpolator="idw", griddata=False, k=10,
                   p_process="mondrian",
                   exponent=list(np.arange(1.0, 15.0, 1.0)),
                   alpha=(0.5, 0.6, 0.8, 0.9, 0.95, 0.98)
                   )
\end{verbatim}
\end{footnotesize}

\begin{figure}[htb]
    \centering
    \includegraphics[width=1\linewidth]{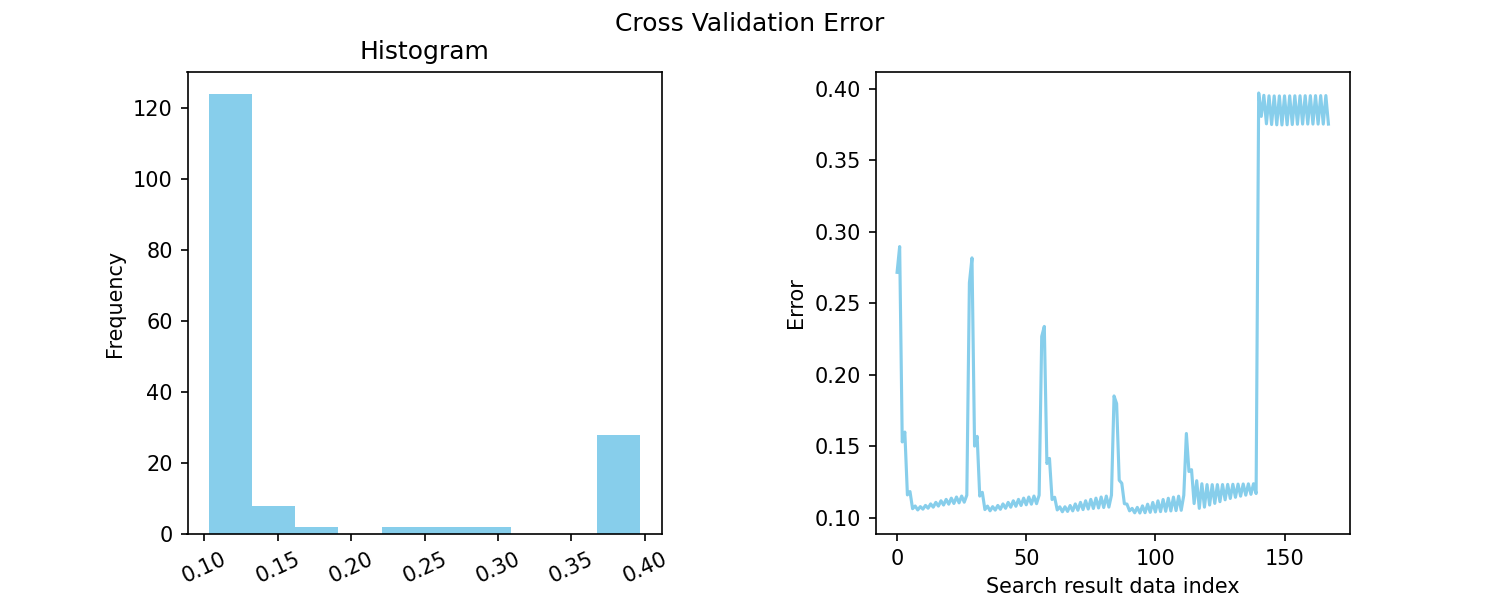}
    \caption{Cross validation error for the ESI-IDW non-gridded estimation parameter grid search.}
    \label{fig:cross_nong_idw}
\end{figure}

Figure \ref{fig:cross_nong_idw} shows the histogram for the cross-validation errors of the 168 search scenarios. In this case, the best-case scenario has the following parameters:

\begin{footnotesize}
    \begin{verbatim}
{'agg_func_name': 'mean', 'cv_error':         
 0.10273662625968455, 'local_interpolator': 'idw', 'exponent': 8.0,     
 'alpha': 0.9, 'result_data_index': 98, 'agg_function': <function mean  
 at 0x14815f060>, 'p_process': 'mondrian'}  
\end{verbatim}
\end{footnotesize}

In this case, the local interpolator that ESI is using is IDW, and we can observe that the chosen exponent is relatively high. Recall that when the exponent is $0$, the local estimate is the simple average among all neighbours, which generates a smoothing effect similar to Kriging. On the other hand, when the exponent tends to infinity, IDW becomes the nearest neighbour estimator, which implies that the estimate acquires a more abrupt change effect.

With these obtained parameters, we then performed the corresponding estimation using the \verb|esi_nongriddata()| function, as shown in Code snipped \ref{listing:15}.

\begin{listing}[H]
\begin{footnotesize}
\begin{verbatim}
result = 
esi_nongriddata(points, values, xi,
                local_interpolator="idw",
                p_process="mondrian",
                n_partitions=500,
                best_params_found=search_result.best_result()
                )
\end{verbatim}
\end{footnotesize}
\caption{Estimation with non-gridded data for ESI-IDW with the best parameters obtained from the previous grid search.}
\label{listing:15}
\end{listing}

Once the prediction was been obtained, the \verb|lf.OperationalErrorLoss()| function was used to generate a custom precision calculation. Specifically, a loss function called ‘Operational Error’ was defined as:

\begin{footnotesize}
    \begin{verbatim}
op_error = lf.OperationalErrorLoss(np.abs(np.nanmin(values) - np.nanmax(values)))
\end{verbatim}
\end{footnotesize}

\begin{figure}[H]
    \centering
    \includegraphics[width=1\linewidth]{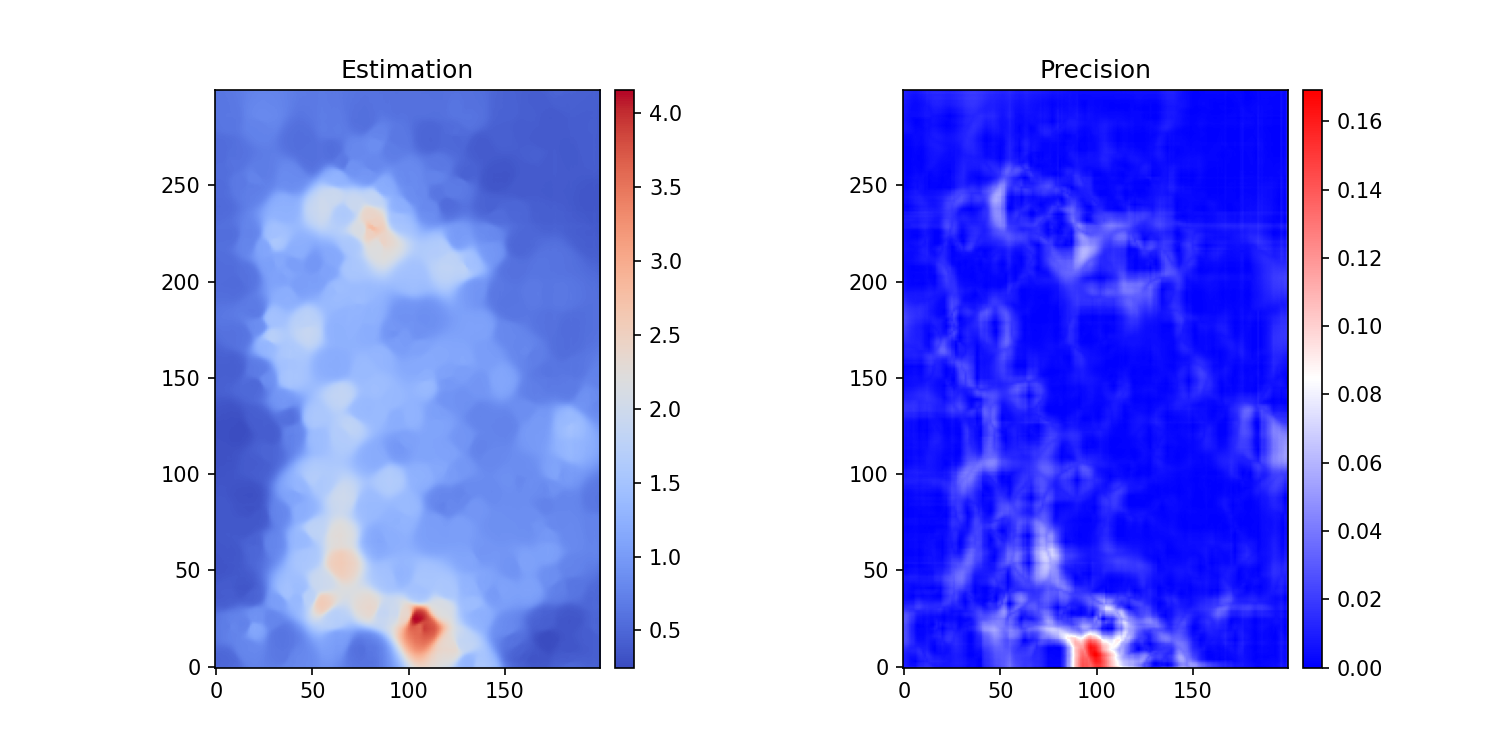}
    \caption{Best parameter non-gridded estimation with ESI-IDW and precision obtained with operational error loss function.}
    \label{fig:nong_idw}
\end{figure}

Figure \ref{fig:nong_idw} shows the resulting estimation and precision using the operational error loss function for the best case of the parameter grid for ESI-IDW. In this Figure, we can observe a more pixelated texture, which is consistent with what was observed above regarding the effect of the abrupt change of a relatively high-value exponent for the local IDW interpolator. Interestingly, this contrasts with the perception, conditioned by the widespread use of Kriging, that a smoothed estimate may be better.

Next, the process of searching for parameters and then estimating using the best set was repeated, but Voronoi was used as the partitioning method.

\begin{figure}[H]
    \centering
    \includegraphics[width=1\linewidth]{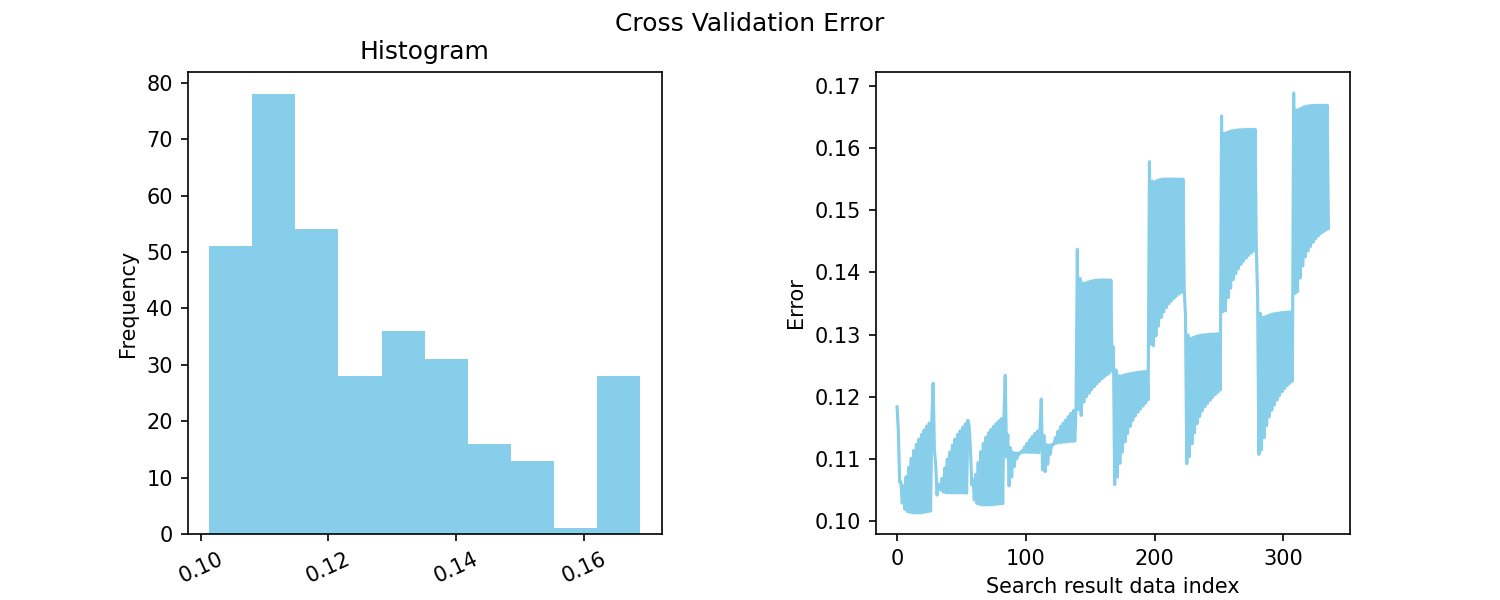}
    \caption{Cross validation error for the ESI-IDW non-gridded estimation parameter grid search, using Voronoi partition method.}
    \label{fig:cross_nong_idw_vor}
\end{figure}

Figure \ref{fig:cross_nong_idw_vor} shows the histogram for the cross-validation errors of the 336 search scenarios. Note that using the Voronoi partitioning method doubles the number of search scenarios due to the \verb|data_cond| parameter, which takes two possible default values (to condition or not condition the partitioning to the samples). In this case, the best-case scenario has the following parameters:

\begin{footnotesize}
    \begin{verbatim}
{'agg_func_name': 'mean', 'cv_error':         
0.10132136752665043, 'local_interpolator': 'idw', 'exponent': 8.0,   
'data_cond': True, 'alpha': 0.5, 'result_data_index': 14,
'agg_function': <function mean at 0x16827b060>, 
'p_process': 'voronoi'}
\end{verbatim}
\end{footnotesize}

Note that in this case, as in the previous case of ESI-IDW parameter search with the Mondrian partitioning method, the exponent for the local interpolator is relatively high. This implies that the estimation should appear to have abruptly changing textures.

\begin{figure}[H]
    \centering
    \includegraphics[width=1\linewidth]{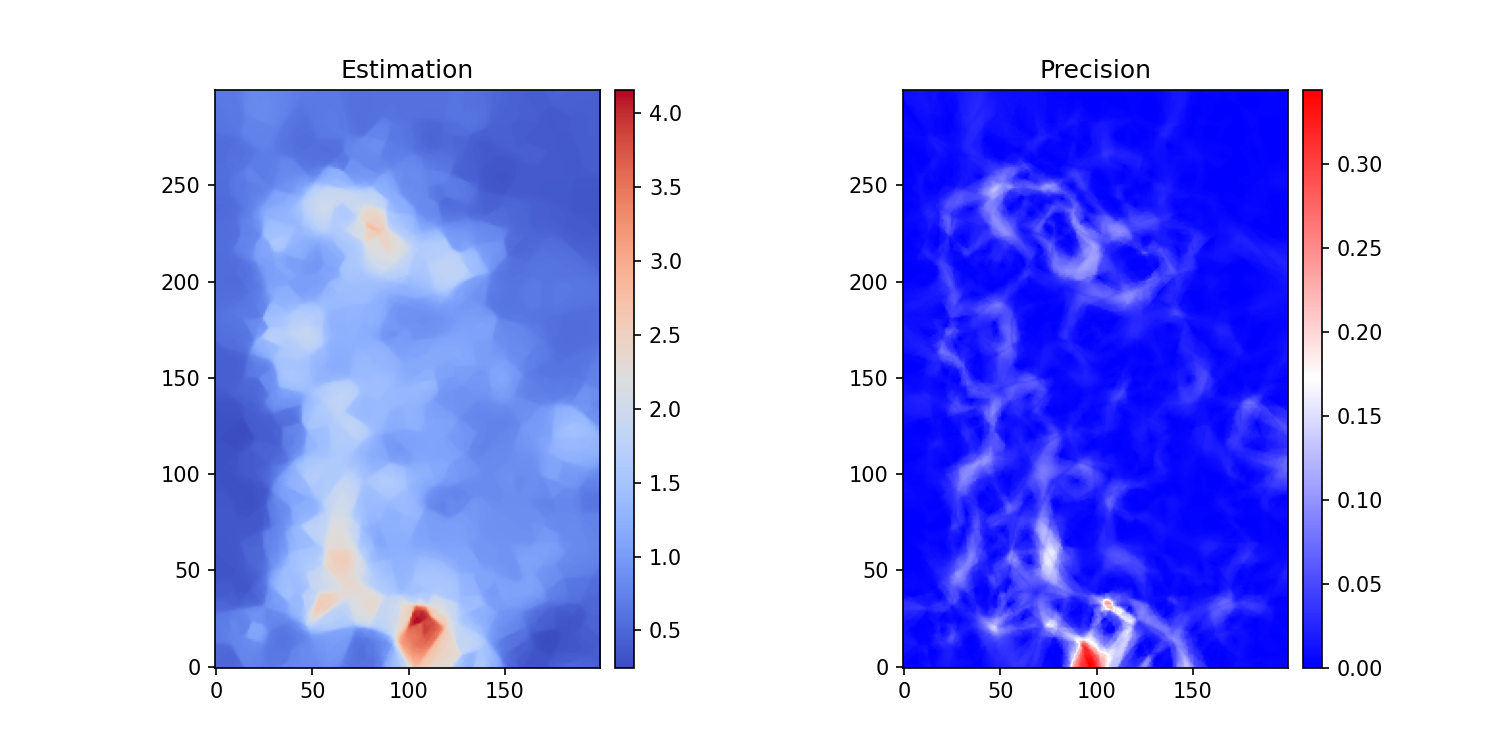}
    \caption{Best parameter non-gridded estimation with ESI-IDW and precision obtained with operational error loss function, this time with Voronoi partition method.}
    \label{fig:nong_idw_vor}
\end{figure}

Next, the corresponding estimation, using the Voronoi partition method, was performed using the \verb|esi_nongriddata()| function. Figure \ref{fig:nong_idw_vor} shows the resulting estimation and precision using the operational error loss function for the best case of the parameter grid for ESI-IDW.

It can indeed be seen that, as in the case of the use of Mondrian as a partitioning method, the texture of the estimation is one of abrupt changes. On the other hand, the precision map is in a notoriously larger range of values, suggesting that, for the evaluated grid of parameters, the optimum for the case of the Voronoi partitioning method achieves a lower accuracy. In this sense, it is important to note that although the same search range was used for the parameter \verb|alpha| in both examples, this parameter has a different implementation in each case -- although it reflects the granularity of the partition in both methods -- which explains this difference. This argument is reinforced by the fact that the resulting value for this parameter ($0.5$) is at the lower end of the range used. 

Analogous to the ESI-IDW case, we present an estimation for the ESI-Kriging case (using Kriging as base interpolator for ESI)\footnote{For the local Kriging interpolator, it is not possible to use the Voronoi partitioning method; this option has not yet been implemented in \texttt{spatialize}}. In this case, the parameter grid includes four models for the omnidirectional covariance function \verb|model|:

\begin{footnotesize}
\begin{verbatim}
search_result = 
esi_hparams_search(points, values, xi,
                   local_interpolator="Kriging", griddata=False, k=10,
                   model=["spherical", "exponential", "cubic", "gaussian"],
                   nugget=[0.5, 1.0],
                   range=[100.0, 500.0, 1000.0],
                   alpha=list(np.flip(np.arange(0.90, 0.95, 0.01))),
                   sill=[0.9, 1.0, 1.1]
                   )
\end{verbatim}
\end{footnotesize}

\begin{figure}[H]
    \centering
    \includegraphics[width=1\linewidth]{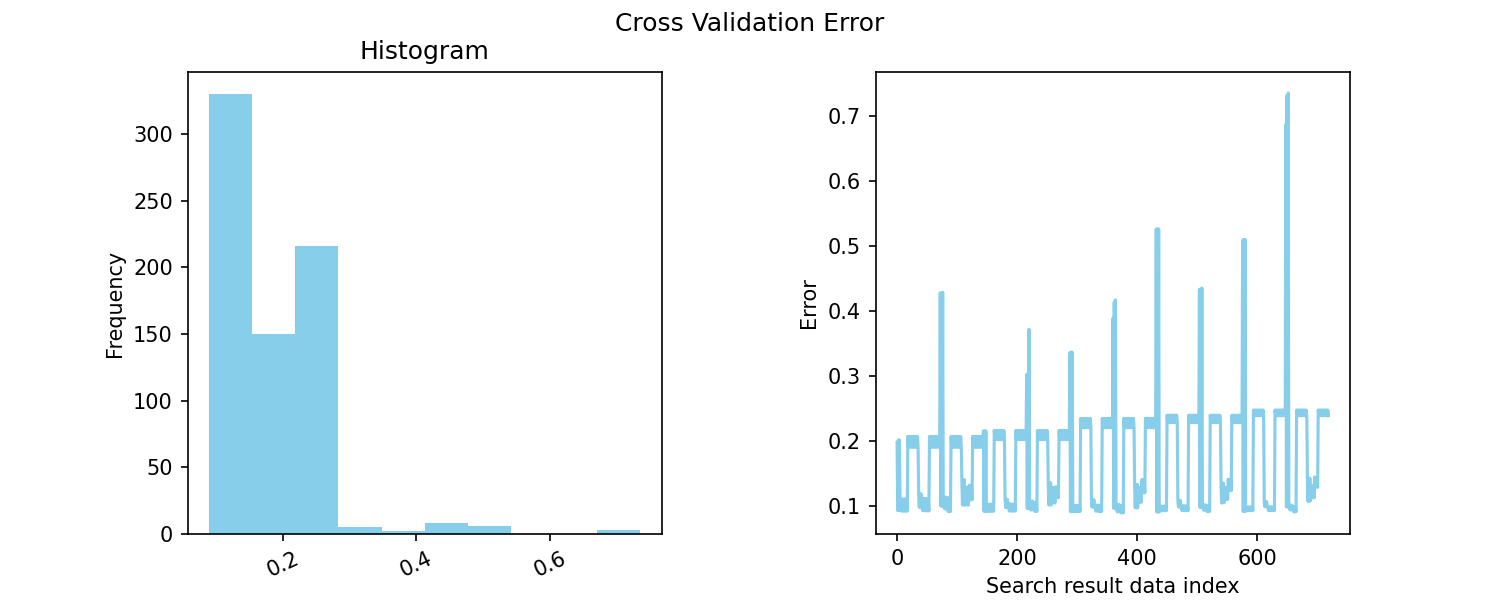}
    \caption{Cross validation error for the ESI-Kriging non-gridded estimation parameter grid search.}
    \label{fig:cross_nong_krig}
\end{figure}

In Figure \ref{fig:cross_nong_krig}, we can see the cross-validation error for the scenarios generated with this parameter search.

\begin{figure}[H]
    \centering
    \includegraphics[width=1\linewidth]{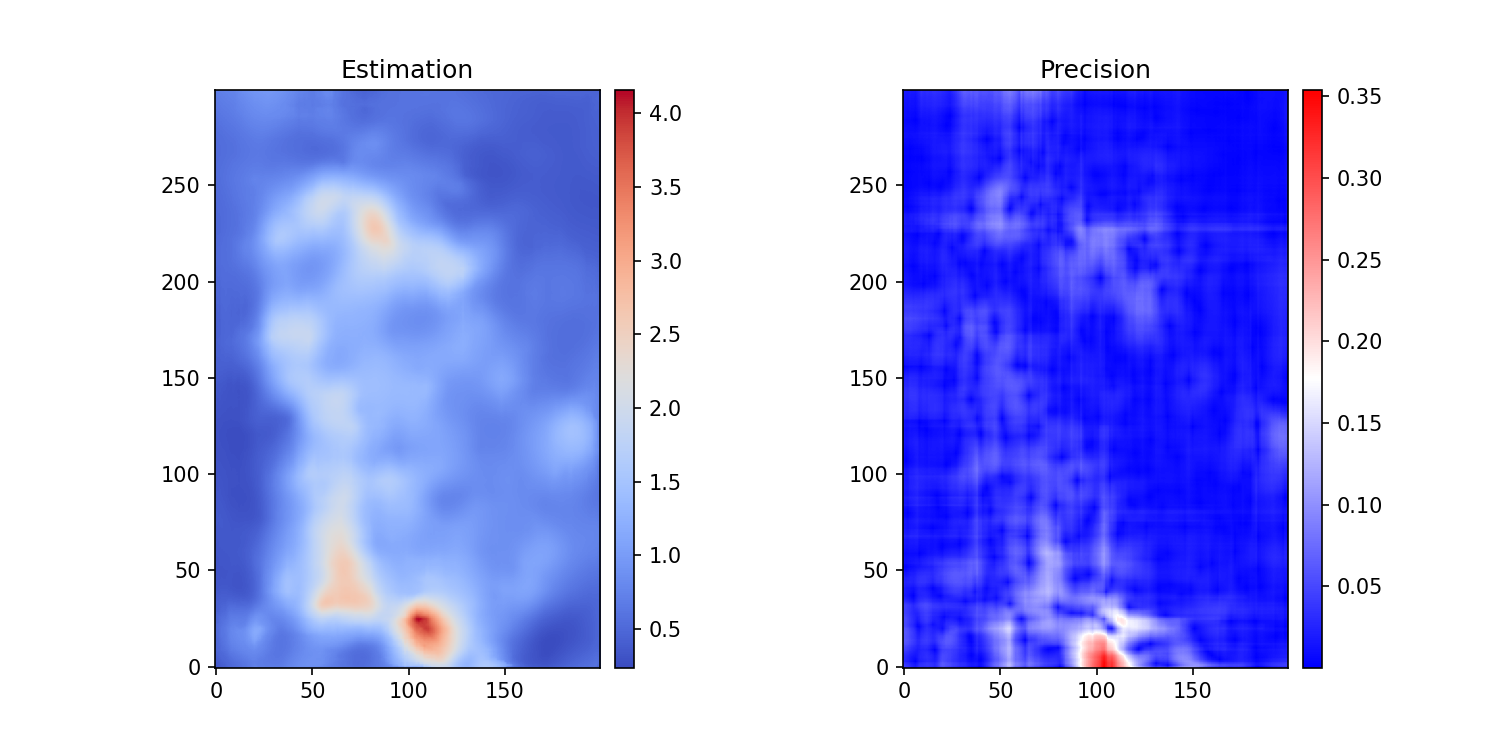}
    \caption{Best parameter non-gridded estimation with ESI-Kriging and precision obtained with operational error loss function.}
    \label{fig:nong_krig}
\end{figure}

Once the best set of parameters was obtained, we generated the corresponding estimates using the \verb|esi_nongriddata()| function. In addition, we generated an accuracy estimate using the operational error in the same way as in the ESI-IDW case. The calculated estimate and precision are shown in Figure \ref{fig:nong_krig}.

Notably, in the ESI-Kriging example in Figure \ref{fig:nong_krig} less smoothing can be noted compared to the Kriging example (Figure  \ref{fig:krig_and_pykrige}), which is generally more credible as an interpolation result.

\subsubsection{3D non-gridded data estimation}

\begin{figure}[H]
    \centering
    \includegraphics[width=1\linewidth]{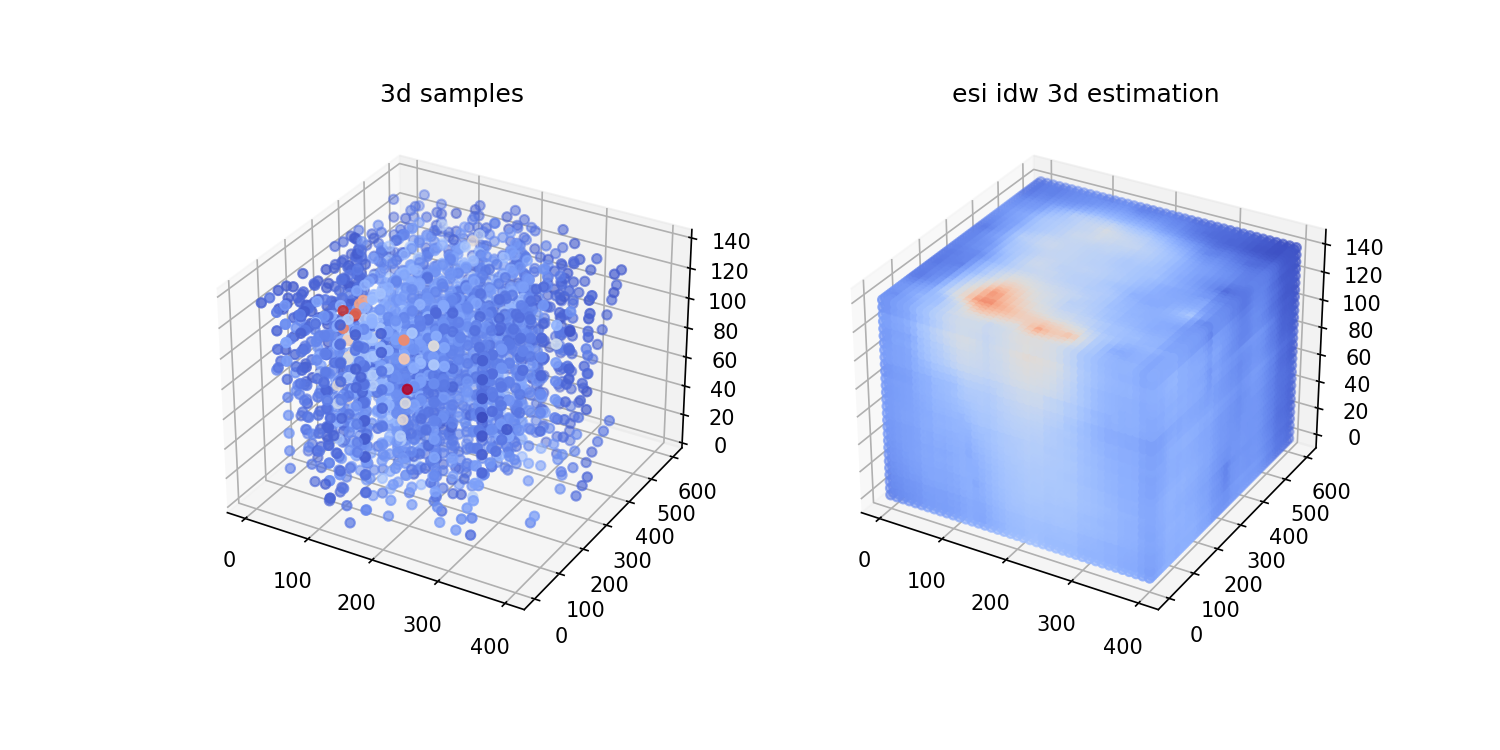}
    \caption{3d data samples and estimation using ESI-IDW}
    \label{fig:3d_data_and_estimation}
\end{figure}

As a complementary analysis, to demonstrate ESI's generalisation capability, the output for a three-dimensional non-gridded data estimation is presented in Figure \ref{fig:3d_data_and_estimation}. The data corresponds to that loaded by using the \verb| load_drill_holes_andes_3D()| function in \verb|spatialize.data|.

\conclusions[Conclusion and future work]  

We have introduced \verb|spatialize|, an open-source library that makes available to the community an efficient implementation of a highly novel geostatistical technique. It aims to provide a general-purpose tool for non-expert geostatistics users that provides automated tools that are on par with expert-use geostatistical tools.

The implemented technique is known as \emph{ensemble spatial interpolation} (ESI). It is essentially data-driven, its philosophy is aligned with computational statistics, and it seeks to bring the power of ensemble learning to geostatistical practice. In this sense, the main strength of \verb|spatialize| is that it requires minimal user intervention, and in cases where it is necessary (such as in the choice of some hyperparameters), it provides help tools that facilitate the process.

The idea is that \verb|spatialize| will be, in the medium term, one of the best open-source geostatistical libraries available in the Python language. To this end, we have a roadmap that includes, as future work:

\begin{itemize}
    \item Add other partition generation processes, such as Mondrian processes with random rotations, to allow for more expressiveness in the set of generated partitions.

    \item Add local interpolators that allow local adaptability in the calculation of ESI samples.

    \item Add local interpolators that allow the use of spatial statistical tools, such as CAR-type models.

    \item Add support for estimation of categorical variables.

    \item On the computational side, add support for the use of GPUs to allow integration with tools such as Google Colab.

    \item Add classical geostatistical functionalities for expert users who require them. This includes a more general Kriging implementation, allowing nested structures, for example.
\end{itemize}





\codedataavailability{
Both the source code for \texttt{spatialize} and the usage examples shown above are available in the spatialize project on Github \citep{spatialize2025} which can be accessed at \href{https://github.com/alges/spatialize}{\url{https://github.com/alges/spatialize}}. In particular, the usage examples are available to be run in the \texttt{examples/scripted\_examples} folder.

The gridded data estimation example (Section \ref{grid_est}) employs the following scripts:
\begin{enumerate}
\setlength{\itemsep}{-1pt}
    \item \texttt{esi\_griddata.py}
    \item \texttt{esi\_griddata\_mondrian\_voronoi.py}
    \item \texttt{esi\_grid\_search.py}
    \item \texttt{idw\_grid\_search.py}
    \item \texttt{idw\_griddata.py}
    \item \texttt{esi\_custom\_precision.py}
\end{enumerate}

The non-gridded data estimation example (Section \ref{non-grid_est}) employs the following scripts:
\begin{enumerate}
\setlength{\itemsep}{-1pt}
    \item \texttt{esi\_nongriddata.py}
    \item \texttt{pykrige\_example.py}
    \item \texttt{esi\_3d\_nongriddata.py}
\end{enumerate}

The data that is used in the scripted examples is part of the \texttt{spatialize} library package and is installed with it in \texttt{spatialize.resources}. Besides, the \texttt{spatialize.data} module contains the functions for loading this data.
} 




\appendix

\section{The Spatialize library: User Manual}
\label{App:user-manual} 

\subsection{Downloading and installation}

The \verb|spatialize| package is compatible with Python 3.8 and later versions, offering support across Linux, macOS, and Windows operating systems. This package is readily accessible through the Python Package Index (PyPI) at \url{https://pypi.org/project/spatialize}. To install \verb|spatialize|, you have two convenient options:
\begin{enumerate}
\item Install from PyPI:

    \verb|pip install spatialize|

\item Install from GitHub:

\verb|pip install -U git+https://github.com/alges/spatialize.git|

\end{enumerate}

These installation methods ensure you can easily integrate \verb|spatialize| into your Python environment, regardless of your preferred source. By using pip, Python’s package installer, you can effortlessly manage dependencies and keep your project up-to-date with the latest \verb|spatialize| features and improvements.

The \verb|spatialize| Python package is built based on some existing dependencies, the most relevant of which are listed as requirements in the package itself. Among them, one can highlight \textbf{NumPy} (\cite{harris2020array}) and \textbf{pandas} (\cite{mckinney2010data}), which are used for data handling (grids and samples), for data import, and to export the estimation results. \textbf{Matplotlib}
(\cite{hunter2007matplotlib}) It is widely used for illustration purposes. \textbf{Scikit-learn} (\cite{scikit-learn}) is used to standardise the parameter search. The next subsections will extend the description of the library.

\subsection{General library description}

The \verb|spatialize| library is built on three levels. The first two constitute the API of this library, which is implemented in Python. The one closest to the user consists of three high-level functions that allow spatial interpolation in two dimensions on gridded and non-gridded data and a hyperparameter search for these cases. On a second level, aggregation functions are provided according to Eq.\ref{ag_func} and functions for estimating precision according to Eq.\ref{eq:precision}. In the latter case, a class is also provided to implement this type of custom function. Finally, at a third level, an efficient implementation of the methods contained in the spatial estimation functions in the C++ language is found. At the third level, \verb|spatialize| is designed and implemented in such a way that it can work in three or more dimensions. In the first two levels, the API provides only two in this first version. 

In addition, \verb|spatialize| includes an efficient implementation of the Inverse Distance Weighting (IDW) spatial estimation method, together with a hyperparameter search function for this tool. A detailed description of these functions is not included in this publication, as they are not part of the ESI method and its alternatives but are only add-ons. However, examples of their use are presented in Section \ref{use_ex}. On the other hand, future versions of the library are planned to integrate an efficient implementation of Kriging, which will include nested structures for the adjustment of the experimental variograms. 

Our library is an efficient tool designed and implemented to work with large datasets. Therefore, the examples provided and described in Section \ref{use_ex} are in the form of scripts, and no Jupyter notebooks have been provided. If the user wants to use this type of environment, the library will work, but it will quickly accumulate rubbish in memory. This is why, in these cases, it is important to work with a limited depth of each tree (use a parameter \verb|alpha| no greater than $0.90$) and a number of trees in the order of $100$. These values will depend on the amount of data, so they are a reasonable estimate for examples such as those presented in our Scripts.  

\subsection{High level API}
\label{sec:high_level_api}

The main objective of \verb|spatialize| is to provide an easy-to-use tool for non-experts in classical geostatistics. Therefore, the high-level API is composed of functions that are very simple to call and hide all the details outlined in the previous sections from the user. In this section, we describe each of these functions in detail, along with their main arguments. 

\subsubsection{General input data format}\label{sec_input_data}

Like any interpolation function, the basic input data are:

\begin{itemize}
    \item \verb|points|: Contains the coordinates of known data points. This is an $N_s \times D$ array, where $N_s$ is the number of data points, and $D$ is the number of dimensions.

    \item \verb|values|: The values associated with each point in points. This must be a 1D array of length $N_s$.

    \item \verb|xi|: If the data are gridded, they correspond to an array of grids of $D$ components, each with the dimensions of one of the grid faces, $d_1 \times d_2 = N_{x^*}$, where $N_{x^*}$ is the total number of unmeasured locations to estimate. Each component of this array represents the coordinate matrix on the corresponding axis, as returned by the functions \verb|numpy.mgrid| in Numpy, or \verb|meshgrid| in Matlab or R.
    
   If the data are not gridded, they are simply the locations at which to evaluate the interpolation. It is then an $N_{x^*} \times D$ array.

    In both cases, $D$ is the dimensionality of each location, which coincides with the dimensionality of the \verb|points|.
\end{itemize}

\subsubsection{ESI estimation of gridded data}
\label{esi_grid}

This is the function used to make an estimate with ESI in the case of sample data and unmeasured locations that are on a grid -- i.e. where the format of \verb|xi| is as detailed in the previous section.

\textbf{Module:}
\begin{footnotesize}
    \begin{verbatim}
    spatialize.gs.esi
\end{verbatim}
\end{footnotesize}

\textbf{Signature:}
\begin{footnotesize}
    \begin{verbatim}
    def esi_griddata(points, values, xi, |\textit{<optional named arguments>}|)
\end{verbatim}
\end{footnotesize}

In the optional named arguments, it is possible to set the local interpolation method to be used (IDW or Kriging), the partition process to generate the random partition set (Voronoi or Mondrian) and other options related to the interpolation methods. The following list contains all these options, with their default values and descriptions.

To define the process of generating random partitions:

\begin{itemize}
    \item \verb|p_process|: Indicates the stochastic process used to generate the random partitions. The options are \verb|"mondrian"|, to use (\ref{eq:TMF}), and \verb|"voronoi"| to use (\ref{eq:TVF}). The default value is \verb|"mondrian"|.
    
    \item \verb|n_partitions|: Number of random partitions ($m$) to be generated. Once the estimation process is completed, each partition corresponds to an estimation scenario. The default value is \verb|500|.
    
    \item \verb|data_cond|: In case of using \verb|p_process="voronoi"|, this parameter indicates whether the random kernel generation process is conditioned by the sample data, as explained in Section \ref{sec:model_training}. The default value is \verb|True|.
    
    \item \verb|alpha|: Sets the coarseness level of each random partition. Corresponds to the $\alpha \in [0,1)$ parameter described in Section \ref{rule_thumb}. As mentioned there, $\alpha = 0$ will generate the coarsest partition, while $\alpha \rightarrow 1$ will generate finer ones. The default value is \verb|0.8|.
\end{itemize}    

To define and configure the local interpolator:

\begin{itemize}    
    \item \verb|local_interpolator|: This is where the function $\textbf{S}_{\mathcal{L}_k}(x^*)$, described in Section \ref{weak_voter}, is defined. In the current version, it can be \verb|"idw"| or \verb|"kriging"| when \verb|p_process="mondrian"| is defined. If \verb|p_process="voronoi"| is defined, it can only be \verb|"idw"|.
\end{itemize}

if \verb|local_interpolator="idw"| is defined, the method described by Equation (\ref{eq_idw_estimation}) is used to implement IDW. 

\begin{equation}\label{eq_idw_estimation}
    IDW(\boldsymbol{x^*}) = \left\{
\begin{array}{cl}
z(\boldsymbol{x}_i)  & \mbox{if } \boldsymbol{x}_i = \boldsymbol{x^*} \\
\frac{\sum_i w_i \cdot z(\boldsymbol{x}_i)}{\sum_i w_i} & \mbox{otherwise}
\end{array}
\right.
\end{equation}

where,

\begin{equation}\label{eq_idw_weights}
w_i =  \frac{1}{d(\boldsymbol{x}_i, \boldsymbol{x^*})^{p}}     
\end{equation}

and $z(\boldsymbol{x}_i)$ are the values of the known samples measured at locations $\boldsymbol{x}_i \in \mathcal{L}_k$, for some element of a particular partition or tree $T_k$. Then, 

\begin{itemize}    
    \item \verb|exponent|: It is the parameter $p$ of Equation \ref{eq_idw_weights}, which defines whether or not to smooth the weight $w_i$ for the sample $\boldsymbol{x}_i$. Note that if $p = 0$, each sample is assigned the same weight in the interpolation (equivalent to the simple mean of the neighbours), and if $p \rightarrow \infty$, the nearest sample is assigned a weight of 1, while all other samples are assigned a weight of 0 (equivalent to a nearest-neighbour interpolation). The default value is \verb|2.0|.

\end{itemize}

If \verb|local_interpolator="kriging"| is defined, a standard ordinary Kriging calculation method is applied, as described in detail in \cite{Chiles2018}. The variogram model used is fixed (i.e. it is not fitted from an experimental variogram) and isotropic, as it is applied to a very small data set (in each cell of each partition). Then, 

\begin{itemize}    
    \item \verb|model|: Indicates the variographic model used for the ordinary Kriging. In \verb|spatialize|, the general variogram has the form: 

    \[
    \gamma(h) = 
    \begin{cases}
    s \cdot \text{m}(h) & \quad 0 \leq \text{m}(h) \leq 1 \\
    0 & \quad \text{m}(h) < 0 \\
    1 & \quad \text{m}(h) > 1
    \end{cases}
    \]

    Where $\text{m}(h) = (1 - n) \cdot (1 - \text{model}(h))$ and $\text{model}(h)$ represents one of the following functions depending on the chosen model:

    \begin{itemize}
    
        \item \verb|"spherical"|:
        \[
        \text{spherical}(h) = 1.5 \frac{h}{r} - 0.5 \left( \frac{h}{r} \right)^3
        \]
    
        \item \verb|"exponential"|:
        \[
        \text{exponential}(h) =  1 - e^{-3 \frac{h}{r}}
        \]
    
        \item \verb|"cubic"|:
        \[
        \text{cubic}(h) = \left( \frac{h}{r} \right)^2 \left( 7 - 8.75 \frac{h}{r} + 3.5 \left( \frac{h}{r} \right)^3 - 0.75 \left( \frac{h}{r} \right)^5 \right)
        \]
    
        \item \verb|"gaussian"|:
        \[
        \text{gaussian}(h) =  1 - e^{-3 \left( \frac{h}{r} \right)^2}
        \]
    
    \end{itemize}
    
    The default model is \verb|"spherical"|. Parameters $n$, $r$ and $s$ must be specified in arguments: 
    \item \verb|nugget| (with default value \verb|0.1|), \verb|range| (with default value \verb|5000.0|) and \verb|sill| (with default value \verb|1.0|), respectively.
\end{itemize}

With the above in place, the aggregation function only remains to be defined for the preliminary estimate returned by this function. Then,

\begin{itemize}    
    \item \verb|agg_function|: Aggregation function $G$ to calculate the estimate at locations \verb|xi| from the ESI samples, as explained in Sections \ref{ESI_sec} and \ref{result_class}. This function can be any of the functions defined in the module \verb|spatialise.gs.esi.aggfunction| or a user-defined custom function (see more details in Section \ref{agg_func}). The default function is \verb|spatialize.gs.esi.aggfunction.mean|.
\end{itemize}

\textbf{Returns:}
\begin{itemize}
    \item (\verb|ESIResult|): As a result, the function returns an object, which is an instance of the class \verb|ESIResult|, containing the preliminary estimate according to the provided arguments. This class provides a set of methods to display aspects of the result, such as the aggregate estimate, the scenarios of the different partitions, or a precision calculation based on some loss function -- for full details, see \ref{result_class}.
\end{itemize}

\subsubsection{ESI estimation of non-gridded data}
\label{esi_nongrid}

This function generates an estimate in ESI space, from a set of sample points (i.e. measured locations), at a set of unmeasured points at arbitrary locations in space. This implies that the \verb|xi| positions must come in the appropriate format, as described in \ref{sec_input_data}.

\textbf{Module:}
\begin{footnotesize}
    \begin{verbatim}
    spatialize.gs.esi
\end{verbatim}
\end{footnotesize}

\textbf{Signature:}
\begin{footnotesize}
    \begin{verbatim}
    def esi_nongriddata(points, values, xi, |\textit{<optional named arguments>}|)
\end{verbatim}
\end{footnotesize}

Except for \verb|xi|, all the arguments of this function are the same as those of the function for gridded data detailed in the previous section (\ref{esi_grid}).

\textbf{Returns:}
\begin{itemize}
    \item (\verb|ESIResult|): As a result, this function also returns an object, which is an instance of the class \verb|ESIResult|, containing the preliminary estimate according to the provided arguments. This class provides a set of methods to display aspects of the result, such as the aggregate estimate, the scenarios of the different partitions, or a precision calculation based on some loss function -- for full details, see \ref{result_class}.
\end{itemize}

\subsubsection{ESIResult class and its methods}
\label{result_class}

Both functions for generating ESI estimates, \verb|esi_nongriddata| and \verb|esi_griddata|, return instances of the \verb|ESIResult| class. An object of this class allows the manipulation of the results in a disaggregated form, the calculation of the final estimate based on different aggregation functions, and the calculation of its precision based on different loss functions. Finally, it allows plotting results in different useful ways.

\begin{footnotesize}
    \begin{verbatim}
    class ESIResult(EstimationResult)
\end{verbatim}
\end{footnotesize}

This class has the following set of methods: 

\begin{itemize}
    \item \verb|esi_samples|: The central concept for dealing with ESI estimation results is the \emph{ESI sample}. In this sense, it should be noted that each random partition delivers an estimate for each of the locations provided in the argument \verb|xi| (for both gridded and non-gridded data). The set of estimates for a particular partition is what in \verb|spatialize| is considered an  \emph{ESI sample}.    

    This method then returns the set of all ESI samples, one for each random partition, calculated for the estimation.
    \begin{itemize}
        \item \textbf{Signature:}
        \begin{footnotesize}
    \begin{verbatim}
def esi_samples()
\end{verbatim}
\end{footnotesize}
        \item \textbf{Returns:}
        \begin{itemize}
            \item (\verb|ndarray|): An array of dimension $N_{x^*} \times m$ ($m$ = \verb|n_partitions| in both function \verb|esi_griddata| and \verb|esi_nongriddata|), for non-gridded data, and of dimension $d_1 \times d_2 \times m$ for gridded data -- remember that, in this case, $d_1 \times d_2 = N_{x^*}$ (see Section \ref{sec_input_data}).
        \end{itemize}
    \end{itemize}
    
    \item \verb|estimation|: Returns the estimated values at locations \verb|xi| by aggregating all ESI samples using the aggregation function provided in the \verb|agg_function| argument (in both functions \verb|esi_griddata| and \verb|esi_nongriddata|). This estimate can be changed using another aggregation function with the \verb|re_estimate| method of this same class.
    \begin{itemize}
        \item \textbf{Signature:}
        \begin{footnotesize}
    \begin{verbatim}
def estimation()
\end{verbatim}
\end{footnotesize}
        \item \textbf{Returns:}
        \begin{itemize}
            \item (\verb|ndarray|): An array of dimension $N_{x^*}$, for non-gridded data, and of dimension $d_1 \times d_2$ for gridded data -- remember that, in this case, $d_1 \times d_2 = N_{x^*}$ (see Section \ref{sec_input_data}). 
        \end{itemize}
    \end{itemize}

    \item \verb|re_estimate|: It recalculates the final estimate based on the aggregation function provided (e.g. by taking the mean of the ESI samples). This method updates the internal estimate and returns the new result. Then, the next time the \verb|estimation| method is called, this is the estimate it will return.
    
    \begin{itemize}
        \item \textbf{Signature:}
        \begin{footnotesize}
    \begin{verbatim}
def re_estimate(agg_function)
\end{verbatim}
\end{footnotesize}
        \item \textbf{Parameters:}
        \begin{itemize}
            \item \verb|agg_function|: Aggregation function $G$ to calculate the estimate at locations \verb|xi| from the ESI samples, as explained in Sections \ref{ESI_sec} and \ref{result_class}. It can be any of the functions defined in the module \verb|spatialise.gs.esi.aggfunction| or a user-defined custom function (see more details in Section \ref{agg_func}). The default function is \verb|spatialize.gs.esi.aggfunction.mean|.
        \end{itemize}
        \item \textbf{Returns:}
        \begin{itemize}
            \item (\verb|ndarray|): An array of dimension $N_{x^*}$, for non-gridded data, and of dimension $d_1 \times d_2$ for gridded data -- remember that, in this case, $d_1 \times d_2 = N_{x^*}$ (see Section \ref{sec_input_data}). 
        \end{itemize}
    \end{itemize}

    \item \verb|precision|: Calculates the precision (or error) between the estimate and the ESI samples using the specified loss function as explained in Section \ref{prec_model}.
    \begin{itemize}
        \item \textbf{Signature:}
        \begin{footnotesize}
    \begin{verbatim}
def precision(loss_function)
\end{verbatim}
\end{footnotesize}
        \item \textbf{Parameters:}
        \begin{itemize}
            \item \verb|loss_function| (\verb|function|, optional): A loss function to calculate the precision, defaulting to \verb|spatialize.gs.esi.lossfunction.mse_loss|. It can be any of the functions defined in the module \verb|spatialise.gs.esi.lossfunction| or a user-defined custom loss function (see more details in Section \ref{loss_func}).
        \end{itemize}
        \item \textbf{Returns:}
        \begin{itemize}
            \item (\verb|ndarray|): An array of dimension $N_{x^*}$, for non-gridded data, and of dimension $d_1 \times d_2$ for gridded data -- remember that, in this case, $d_1 \times d_2 = N_{x^*}$ (see Section \ref{sec_input_data}).
        \end{itemize}
    \end{itemize}
    
    \item \verb|precision_cube|: It applies a loss (error) function to each ESI sample with respect to the current estimate. The difference with the \verb|precision| method is that it does not aggregate the result over the total calculated losses, returning the total data ``cube" whose dimensions are the same as the ESI samples cube.
    \begin{itemize}
        \item \textbf{Signature:}
        \begin{footnotesize}
    \begin{verbatim}
def precision_cube(loss_function=mse_cube)
\end{verbatim}
\end{footnotesize}
        \item \textbf{Parameters:}
        \begin{itemize}
            \item \verb|loss_function| (\verb|function|, optional): A loss function to calculate the precision cube, defaulting to \verb|spatialise.gs.esi.lossfunction.mse_cube|. It can be any of the functions contained in module \verb|spatialise.gs.esi.lossfunction| (or a custom user-defined) whose aggregation function is the \verb|identity| function (which produces a data cube as a result).
        \end{itemize}
        \item \textbf{Returns:}
        \begin{itemize}
             \item (\verb|ndarray|): An array of dimension $N_{x^*} \times m$ ($m$ = \verb|n_partitions| in both function \verb|esi_griddata| and \verb|esi_nongriddata|), for non-gridded data, and of dimension $d_1 \times d_2 \times m$ for gridded data -- remember that, in this case, $d_1 \times d_2 = N_{x^*}$ (see Section \ref{sec_input_data}).
        \end{itemize}
    \end{itemize}

    \item \verb|plot_estimation|: Plots the estimation using \verb|matplotlib|.
    \begin{itemize}
        \item \textbf{Signature:}
        \begin{footnotesize}
    \begin{verbatim}
def plot_estimation(ax, w, h, **figargs)
\end{verbatim}
\end{footnotesize}
        \item \textbf{Parameters:}
        \begin{itemize}
            \item \verb|ax| (\verb|matplotlib.axes.Axes|, optional): The \verb|Axes| object to render the plot on. If \verb|None|, a new \verb|Axes| object is created.
            \item \verb|w| (\verb|int|, optional): The width of the image (if the data is reshaped).
            \item \verb|h| (\verb|int|, optional): The height of the image (if the data is reshaped).
            \item \verb|**figargs| (optional): Additional keyword arguments passed to the figure creation (e.g., DPI, figure size).
        \end{itemize}
    \end{itemize}
    
    \item \verb|plot_precision|: Plots the precision using \verb|matplotlib|. If the precision has not been computed yet, it calls the \verb|self.precision()| method to calculate it.
    \begin{itemize}
        \item \textbf{Signature:}
        \begin{footnotesize}
    \begin{verbatim}
def plot_precision(ax, w, h, **figargs)
\end{verbatim}
\end{footnotesize}
        \item \textbf{Parameters:}
        \begin{itemize}
            \item \verb|ax| (\verb|matplotlib.axes.Axes|, optional): The \verb|Axes| object where the precision plot will be rendered. If \verb|None|, a new \verb|Axes| object will be created.
            \item \verb|w| (\verb|int|, optional): The width of the image (if the data is reshaped).
            \item \verb|h| (\verb|int|, optional): The height of the image (if the data is reshaped).
            \item \verb|**figargs| (optional): Additional keyword arguments passed to the figure creation (e.g., DPI, figure size).
        \end{itemize}
    \end{itemize}
    
    \item \verb|quick_plot|: Creates a quick, side-by-side plot of the \verb|estimation| and \verb|precision| using \verb|matplotlib|. It uses \verb|self.plot_estimation()| and \verb|self.plot_precision()| to render the plots. The figure is returned for further use or display.
    \begin{itemize}
        \item \textbf{Signature:}
        \begin{footnotesize}
    \begin{verbatim}
def quick_plot(w, h, **figargs)
\end{verbatim}
\end{footnotesize}
        \item \textbf{Parameters:}
        \begin{itemize}
            \item \verb|w| (\verb|int|, optional): The width of the image (if the data is reshaped).
            \item \verb|h| (\verb|int|, optional): The height of the image (if the data is reshaped).
            \item \verb|**figargs| (optional): Additional keyword arguments passed to the figure creation (e.g., DPI, figure size).
        \end{itemize}
    \end{itemize}
\end{itemize}

\subsubsection{ESI hyperparameter search}
\label{hyper_p_search}

Apart from providing a general-purpose geostatistical tool (i.e. for non-experts in geostatistics), \verb|spatialize| also aims to make the process as automatic as possible. Thus, the function presented in this section allows a grid search of the best parameters for an estimate. This is done by using parameterizable cross-validation on training and test sets, either with $k$-fold, with random subsets, or \emph{leave-one-out} for exhaustive estimation. 

\textbf{Module:}
\begin{footnotesize}
    \begin{verbatim}
    spatialize.gs.esi
\end{verbatim}
\end{footnotesize}

\textbf{Signature:}
\begin{footnotesize}
    \begin{verbatim}
    def esi_hparams_search(points, values, xi, |\textit{<optional named arguments>}|)
\end{verbatim}
\end{footnotesize}

The mode of use is very similar to that of \verb|esi_griddata| and \verb|esi_nongriddata|. The following parameters are fixed and have the same meaning as in these functions (Sections \ref{sec_input_data}, \ref{esi_grid} and \ref{esi_nongrid}):

\begin{itemize}
    \item \verb|points|, \verb|values|, \verb|xi|
    \item \verb|p_process|
    \item \verb|local_interpolator|
\end{itemize}

In addition to the latter, the following must also be defined:

\begin{itemize}
    \item \verb|k|: Number of subsets for the $k$-fold round. If \verb|k|=$N_{x^*}$ or \verb|k=-1| a \emph{leave-one-out} round is run. The default value is \verb|10|.
    \item \verb|griddata|: If \verb|True|, it is to indicate that the estimation is on a grid (\verb|xi|). The default value is \verb|False|.
\end{itemize}

The grid search parameters must be entered as a set of options in the domain of the argument to be searched -- each argument has the same description and scope as in \verb|esi_griddata| and \verb|esi_nongriddata|. Then:

\begin{itemize}
    \item \verb|data_cond (list)|
    \begin{itemize}
        \item Valid only when \verb|p_process="voronoi"|.
        \item \textbf{Usage example}: \verb|data_cond=[True, False]|
    \end{itemize}
    \item \verb|n_partitions (list)|
    \begin{itemize}
        \item \textbf{Usage example}: \verb|n_partitions=[150, 100, 50]|
    \end{itemize}
    \item \verb|alpha (list)|
    \begin{itemize}
        \item \textbf{Usage example}: \verb|alpha=list(np.flip(np.arange(0.70, 0.90, 0.01)))|
    \end{itemize}
\end{itemize}

if \verb|local_interpolator="idw"| is defined,

\begin{itemize}
    \item \verb|exponent (list)|
    \begin{itemize}
        \item \textbf{Usage example}: \verb|exponent=list(np.arange(1.0, 15.0, 1.0))|
    \end{itemize}
\end{itemize}

if \verb|local_interpolator="kriging"| is defined,

\begin{itemize}
    \item \verb|model (list)|
    \begin{itemize}
        \item \textbf{Usage example}: \verb|model=["spherical", "exponential", "cubic", "gaussian"]|
    \end{itemize}
    \item \verb|nugget (list)|: 
    \begin{itemize}
        \item \textbf{Usage example}: \verb|nugget=[0.0, 0.5, 1.0]|
    \end{itemize}
    \item \verb|range (list)| 
    \begin{itemize}
        \item \textbf{Usage example}: \verb|range=[10.0, 50.0, 100.0, 200.0]|
    \end{itemize}
    \item \verb|sill (list)|
    \begin{itemize}
        \item \textbf{Usage example}: \verb|sill=[0.9, 1.0, 1.1]|
    \end{itemize}
\end{itemize}

Finally, one can also search for an optimal aggregation function with:

\begin{itemize}
    \item \verb|agg_function (dict)|:
    \begin{itemize}
        \item This dictionary can include any of the functions defined in the \\ \verb|spatialise.gs.esi.aggfunction| module or any user-defined custom functions (see more details in the \ref{agg_func} section).
        \item \textbf{Usage example}: \verb|agg_function={"mean": af.mean, "median": af.median}| (if \verb|af| has been imported with \verb|spatialize.gs.esi.aggfunction as af|).
    \end{itemize}
\end{itemize}

\textbf{Returns:}
\begin{itemize}
    \item (\verb|ESIGridSearchResult|) This function returns an instance of the class \verb|ESIGridSearchResult|, detailed below (Section \ref{search_result_class}). In addition, an example of the use of these tools is shown in Section \ref{grid_est_param}.
\end{itemize}
 
\subsubsection{ESIGridSearchResult class and its methods}
\label{search_result_class}

This class contains the results returned by the parameter search function \verb|esi_hparams_search()|, described in Section \ref{hyper_p_search}.

\begin{footnotesize}
    \begin{verbatim}
    class ESIGridSearchResult(GridSearchResult)
\end{verbatim}
\end{footnotesize}

It has two methods:

\begin{itemize}
    \item \verb|best_result|: Constructs a dictionary with the parameters to make the estimate (gridded or non-gridded) corresponding to the smallest cross-validation error in the hyperparameter search made by the \verb|esi_hparams_search()| function. This dictionary is not intended to be manipulated by the user but to be passed directly to the \verb|esi_griddata| and \verb|esi_nongriddata| functions. Thus, a typical call to this method would be:

    \begin{footnotesize}
    \begin{verbatim}
search_result = esi_hparams_search(points, values, (grid_x, grid_y), 
                                   griddata=True, 
                                   <...>)
result = esi_griddata(points, values, 
                      (grid_x, grid_y),
                      best_params_found=search_result.best_result())
    \end{verbatim}
\end{footnotesize}
 
    \begin{itemize}
        \item \textbf{Signature:}
        \begin{footnotesize}
    \begin{verbatim}
def best_result()
\end{verbatim}
\end{footnotesize}
        \item \textbf{Returns:}
        \begin{itemize}
            \item (\verb|dict|): The dictionary with the optimal values found in the cross-validation.
        \end{itemize}
    \end{itemize}

    \item \verb|plot_cv_error|: It shows a graph of the cross-validation errors of the hyperparameter search process. The graph has two components: the first is the error histogram, and the second is the error level for each of the estimation scenarios generated by the gridded parameter search.
    \begin{itemize}
        \item \textbf{Signature:}
        \begin{footnotesize}
    \begin{verbatim}
def plot_cv_error()
\end{verbatim}
\end{footnotesize}
\end{itemize}
\end{itemize}

\subsection{Low level API}
\label{sec:low_level_api}

The high-level API is intended for users who are not experts in either geostatistics or computer programming. In this section, we describe some features of \verb|spatialize| that require some Python programming skills and are intended to make the most of the output of ESI estimations in terms of analysis.

\subsubsection{Aggregation functions}
\label{agg_func}

The aggregation functions $G$, as presented in Section \ref{ESI_sec}, are contained in the module:

\begin{footnotesize}
    \begin{verbatim}
    spatialize.gs.esi.aggfunction
\end{verbatim}
\end{footnotesize}

Their main use is to aggregate the ESI samples generated by the ESI estimation process. However, they are also useful for aggregating the samples generated by the precision calculation between an aggregated estimate and the ESI samples of the estimate. The latter is discussed in more detail in the section \ref{loss_func}.

In practice, any function with the following signature can be considered as an aggregation function in \verb|spatialize|:

\begin{footnotesize}
\begin{verbatim}
def f(samples):
    ...
\end{verbatim}
\end{footnotesize}

The \verb|samples| argument of \verb|f| is always considered to be non-gridded -- that is, an array of dimension $N_{x^*} \times m$ ($m$ = \verb|n_partitions| in both \verb|esi_griddata| and \verb|esi_nongriddata|). When the estimates are gridded, \verb|spatialize| flattens them to treat them internally as non-gridded and reshapes them on return. In this context, the expected behaviour of \verb|f| is to return an array of dimension $N_{x^*}$ with the aggregate on the second axis of the samples, i.e. aggregating the $m$ \verb|samples|. For example, the code for the predefined \verb|mean| aggregation function in \verb|spatialize| is:

\begin{footnotesize}
\begin{verbatim}
import numpy as np

def mean(samples):
    return np.nanmean(samples, axis=1)
\end{verbatim}
\end{footnotesize}

The standard aggregation functions included in the library are:

\begin{itemize}

    \item \verb|mean|: Calculates the mean of a set of samples.
    \begin{itemize}
        \item \textbf{Signature:} 
        \begin{footnotesize}
    \begin{verbatim}
def mean(samples)
        \end{verbatim}
\end{footnotesize}
        \item \textbf{Returns:} The arithmetic mean of the provided samples.
    \end{itemize}
    
    \item \verb|median|: Calculates the median of a set of samples.
    \begin{itemize}
        \item \textbf{Signature:} 
        \begin{footnotesize}
    \begin{verbatim}
def median(samples)
        \end{verbatim}
\end{footnotesize}
        \item \textbf{Returns:} The median value of the provided samples.
    \end{itemize}
    
    \item \verb|MAP|: Calculates the Maximum A Posteriori (MAP) estimate.
    \begin{itemize}
        \item \textbf{Signature:} 
        \begin{footnotesize}
    \begin{verbatim}
def MAP(samples)
        \end{verbatim}
\end{footnotesize}
        \item \textbf{Returns:} In the current version, the mode of the empirical distribution is returned.
    \end{itemize}
    
    \item \verb|class Percentile|: This is a class for creating functions (callable instances) that belong to a family of functions indexed by a given \verb|percentile|. 
    \begin{itemize}
        \item \textbf{Constructor signature:} 
        \begin{footnotesize}
    \begin{verbatim}
class Percentile(percentile)
        \end{verbatim}
\end{footnotesize}
        \item \textbf{Returns:} When the constructor of the class is called, it returns a function that calculates the \verb|percentile|, passed as an argument, over the samples. The returned function has the general signature of an aggregation function. For example, the code to obtain an aggregation function \verb|p75| that calculates the 75th percentile is:

       \begin{footnotesize}
    \begin{verbatim}
p75 = Percentile(75)           
       \end{verbatim}
\end{footnotesize}   
       
       The function \verb|p75|, for example, can be used to make a call such as 

       \begin{footnotesize}
    \begin{verbatim}
result = esi_griddata(..., 
                      agg_function=p75,
                      ...)      
       \end{verbatim}
\end{footnotesize}  
    \end{itemize}
    
    \item \verb|class WeightedAverage|: This is a class for creating functions (callable instances) that belong to a family of functions indexed by a given set of \verb|weights|. 
        \begin{itemize}
        \item \textbf{Constructor signature:} 
        \begin{footnotesize}
    \begin{verbatim}
class WeightedAverage(weights, normalize, force_resample)
        \end{verbatim}
\end{footnotesize}
        \item \textbf{Returns:} When the class constructor is called, it returns a function, say \verb|ws|, which calculates a weighted average of the samples provided. By default, when the \verb|weights| are not provided as an argument in the constructor, they are generated randomly, drawn from a Dirichlet distribution. If \verb|force_resample=True| is also specified (this is the default), the weights will be generated each time \verb|ws| is called. Finally, if \verb|normalize=True| is defined (default value is \verb|False|), the result returned by \verb|ws| is normalised to the range of the samples to avoid excessive smoothing for some set of weights.

        For example, when used in the following way, \verb|ws| is called only once:
       \begin{footnotesize}
    \begin{verbatim}
ws = WeightedAverage()
result = esi_griddata(..., 
                      agg_function=ws,
                      ...)
        \end{verbatim}
\end{footnotesize} 

        But in the following lines of code, \verb|e1| and \verb|e2| will be different estimates because the weights will be different for each call:        

        \begin{footnotesize}
    \begin{verbatim}
result.re_estimate(ws)
e1 = result.estimation()

result.re_estimate(ws)
e2 = result.estimation()
        \end{verbatim}
\end{footnotesize}

    \end{itemize}
    
    \item \verb|bilateral_filter|: Non-linear filter acting on ESI samples as an aggregation function. It works in both spatial and sample dimensions. Its effect is that in the final estimation, it reduces noise while preserving the areas where there are abrupt changes (edges or high-frequency areas) in the variable to be estimated (\cite{Tomasi1998}). 
    \begin{itemize}
        \item \textbf{Signature:} 
        \begin{footnotesize}
    \begin{verbatim}
def bilateral_filter(samples)
        \end{verbatim}
\end{footnotesize}
        \item \textbf{Returns:} A filtered version of the input cube, with reduced noise and preserved edges.
    \end{itemize}

    \item \verb|identity|: Its main function is to be passed as an aggregation function in the definition of ``cube" loss functions, where only the loss (error) between the ESI samples and the estimate is calculated, but the result is not aggregated. This leaves the possibility to explore more complex aggregation functions to calculate precision -- more details in Section \ref{loss_func}.
    \begin{itemize}
        \item \textbf{Signature:} 
        \begin{footnotesize}
    \begin{verbatim}
def identity(samples)
        \end{verbatim}
\end{footnotesize}
        \item \textbf{Returns:} Returns the input data as-is (identity function).
    \end{itemize}
    
\end{itemize}

\subsubsection{Loss functions}
\label{loss_func}

We will now review the module content.

\begin{lstlisting}[language=Python, mathescape=true, numbers=none]
    spatialize.gs.esi.lossfunction
\end{lstlisting}

Before going into the functions themselves, as mentioned above, \verb|spatialize| provides a powerful framework to easily build any uncertainty quantification scheme based on the precision model expressed in Equation (\ref{eq:precision}). It is useful to review that model here to identify its components and to understand how it is expressed in the \verb|spatialize| idiom. Thus, in Equation (\ref{eq:precision}), $\mathbb{E}$ represents an aggregation function, $\mathbb{L}$ a loss (error) function, $e^*$ is the estimate and $\{x_k^*\}_m$ are the ESI samples. With these components, the precision function $p^*$ can be implemented in \verb|spatialize| with the following idiomatic structure:

\begin{footnotesize}
\begin{verbatim}
from spatialize.gs.esi.lossfunction import loss

@loss(E)
def p(x, y):
    return L(x, y)
\end{verbatim}
\end{footnotesize}

Note that in this structure, neither $e^*$ nor $\{x_k^*\}_m$ appears explicitly. This is because the \verb|@loss| decorator internally transforms the function \verb|p| to have the signature:

\begin{footnotesize}
\begin{verbatim}
def p(estimation, esi_samples):
    ...
\end{verbatim}
\end{footnotesize}

The aim of this is that the design effort is put into the development of the function \verb|L|, leaving \verb|spatialize| to handle the internal processing of the more complex behaviour of \verb|p|. In this way, for example, the precision function of Equation (\ref{eq:interpolator_error}) can be implemented very easily as:

\begin{footnotesize}
\begin{verbatim}
from spatialize.gs.esi.lossfunction import loss
from spatialize.gs.esi.agg_function import mean

@loss(mean)
def p_E(x, y):
    return (x - y) ** 2
\end{verbatim}
\end{footnotesize}

In other words, when decorating the function \verb|p_E| with \verb|@loss|, giving it an aggregation function (\verb|mean|, in this case), it becomes the implementation of a precision model. Once this has happened, it can be passed as an argument when calculating the precision of the result of an ESI estimation, as shown below:

\begin{footnotesize}
\begin{verbatim}
result = esi_griddata(...)
p = result.precision(p_E)
\end{verbatim}
\end{footnotesize}

Two very common loss functions are included in this module:

\begin{itemize}
    \item \verb|mse_loss|: Implements the precision model given by Equation (\ref{eq:interpolator_error}).

    \item \verb|mae_loss|: Implements the precision model given by
        \begin{equation}
           p_{\mathbb{E}}^* = \dfrac{1}{m}\sum_{k=1}^m |x_k^* - e_{\mathbb{E}}^*|
        \end{equation}   
\end{itemize}

An example of a call for both is

\begin{footnotesize}
\begin{verbatim}
from spatialize.gs.esi.lossfunction import mse_loss

result = esi_griddata(...)
p = result.precision(mse_loss)
\end{verbatim}
\end{footnotesize}

Where \verb|p| is an array of dimension $d_1 \times d_2=N_{x^*}$ in the case of gridded data, or simply an array of dimension $N_{x^*}$, for the non-gridded case.

In addition, versions without aggregation are included, i.e. where the aggregation function is \verb|identity|. These are:

\begin{itemize}
    \item \verb|mse_cube|
    \item \verb|mse_cube|
\end{itemize}

Recall that these last two are to leave the possibility to explore more complex aggregation functions to calculate accuracy. Both can be passed as arguments when calculating the precision, as shown below: 

\begin{footnotesize}
\begin{verbatim}
from spatialize.gs.esi.lossfunction import mse_cube

result = esi_griddata(...)
p = result.precision(mse_cube)
\end{verbatim}
\end{footnotesize}

Just note that \verb|p| is no longer an array of dimensions $d_1 \times d_2=N_{x^*}$, but an array of dimension $d_1 \times d_2 \times m$ ($m$ = \verb|n_partitions| in both \verb|esi_griddata| and \verb|esi_nongriddata|) in the case of gridded data (or an array of dimension $N_{x^*} \times m$, for the non-gridded case), where $m$ is the number of ESI samples.

Finally, the module also includes:

\begin{itemize}
    \item \verb|class OperationalErrorLoss|: This is a class for creating functions (callable instances) that belong to a family of functions indexed by a given \emph{dynamic range}. 
    \begin{itemize}
        \item \textbf{Constructor signature:} 
        \begin{footnotesize}
    \begin{verbatim}
class OperationalErrorLoss(dyn_range, use_cube)
        \end{verbatim}
\end{footnotesize}
        \item \textbf{Returns:} When the constructor of the class is called, it returns a function that implements the following precision model: 

        \begin{equation}
           p_{\mathbb{E}}^* = \dfrac{1}{m \cdot d_r}\sum_{k=1}^m |x_k^* - e_{\mathbb{E}}^*|
        \end{equation}          

        Where $d_r$ is an expected dynamic range for output errors, passed in the argument \verb|dyn_range| (default value is \verb|None|). If \verb|dyn_range| is not passed in the argument, the dynamic range of the estimate ($e_{\mathbb{E}}^*$) is used. If \verb|use_cube=True| (default value \verb|False|), then the returned function is decorated with the \verb|identity|, having the same output behaviour as \verb|mse_cube| or \verb|mae_cube|. 

        An example of a call is as follows:

       \begin{footnotesize}
    \begin{verbatim}
from spatialize.gs.esi.lossfunction import OperationalErrorLoss

result = esi_griddata(...)

op_error_loss = OperationalErrorLoss(100)

p = result.precision(op_error_loss)    
       \end{verbatim}
\end{footnotesize}  
    \end{itemize}
\end{itemize}

\noappendix       







\section{}
\authorcontribution{
Research conceptualisation, paper preparation, and analysis were performed by AFE, AE and FN. The ESI framework was first developed by FG, FN and AFE. FN, AFE, AE, FG, MJV, and JFS contributed to the paper edits and technical review.
} 

\competinginterests{The contact author has declared that none of the authors has any competing interests.} 


\begin{acknowledgements}
This work was supported in part by the Chilean National Agency for Research and Development (ANID) under Project FONDEF IT23I013, FONDEF TA24I10053 and ANID/BASAL AFB230001. We acknowledge the computational resources provided, the useful insights, and the support provided by the whole ALGES team.
\end{acknowledgements}







\bibliographystyle{copernicus}
\bibliography{references.bib}

\end{document}